\documentclass[times,twocolumn]{aastex62}

\usepackage{graphicx}
\usepackage{subfigure}
\usepackage{natbib}
\usepackage{color}
\usepackage{tabularx}
\usepackage{bm}
\usepackage{threeparttable}

\newcommand{\tess}{{\it TESS}}

\newcommand{\lco}{{\it LCO}}
\newcommand{\tar}{LHS 1815}
\newcommand{\gaia}{{\it Gaia}}
\newcommand{\jwst}{{\it JWST}}

\begin{document}
\title{{\Large LHS 1815b: The First Thick-Disk Planet Detected By TESS}}
\author[0000-0002-4503-9705]{Tianjun~Gan}
\affil{Department of Astronomy and Tsinghua Centre for Astrophysics, Tsinghua University, Beijing 100084, China}

\author[0000-0002-1836-3120]{Avi Shporer}
\affiliation{Department of Physics and Kavli Institute for Astrophysics and Space Research, Massachusetts Institute of Technology, Cambridge, MA 02139, USA}

\author[0000-0002-4881-3620]{John H. Livingston}
\affiliation{Department of Astronomy, The University of Tokyo, 7-3-1 Hongo, Bunkyo-ku, Tokyo 113-0033, Japan}

\author[0000-0001-6588-9574]{Karen A.\ Collins}
\affiliation{Center for Astrophysics ${\rm \mid}$ Harvard {\rm \&} Smithsonian, 60 Garden Street, Cambridge, MA 02138, USA}

\author[0000-0001-8317-2788]{Shude~Mao}
\affil{Department of Astronomy and Tsinghua Centre for Astrophysics, Tsinghua University, Beijing 100084, China}
\affil{National Astronomical Observatories, Chinese Academy of Sciences, 20A Datun Road, Chaoyang District, Beijing 100012, China}

\author[0000-0001-5371-3432]{Alessandro A. Trani}
\affiliation{Department of Astronomy, The University of Tokyo, 7-3-1 Hongo, Bunkyo-ku, Tokyo 113-0033, Japan}

\author[0000-0001-8627-9628]{Davide Gandolfi}
\affiliation{Dipartimento di Fisica, Universit\`a degli Studi di Torino, via Pietro Giuria 1, I-10125, Torino, Italy}

\author[0000-0003-3618-7535]{Teruyuki Hirano}
\affiliation{Department of Earth and Planetary Sciences, Tokyo Institute of Technology, 2-12-1 Ookayama, Meguro-ku, Tokyo 152-8551, Japan}

\author[0000-0002-4671-2957]{Rafael Luque}
\affiliation{Instituto de Astrof\'isica de Canarias (IAC), 38205 La Laguna, Tenerife, Spain}
\affiliation{Departamento de Astrof\'isica, Universidad de La Laguna (ULL), 38206, La Laguna, Tenerife, Spain}

\author[0000-0002-3481-9052]{Keivan G.\ Stassun}
\affiliation{Department of Physics and Astronomy, Vanderbilt University, 6301 Stevenson Center Ln., Nashville, TN 37235, USA}
\affiliation{Department of Physics, Fisk University, 1000 17th Avenue North, Nashville, TN 37208, USA}

\author[0000-0002-0619-7639]{Carl Ziegler}
\affiliation{Dunlap Institute for Astronomy and Astrophysics, University of Toronto, 50 St. George Street, Toronto, Ontario M5S 3H4, Canada}

\author[0000-0002-2532-2853]{Steve~B.~Howell}
\affiliation{NASA Ames Research Center, Moffett Field, CA 94035, USA}

\author{Coel~Hellier}
\affiliation{Astrophysics Group, Keele University, Staffordshire, ST5 5BG, UK}

\author{Jonathan M.~Irwin}
\affiliation{Center for Astrophysics ${\rm \mid}$ Harvard {\rm \&} Smithsonian, 60 Garden Street, Cambridge, MA 02138, USA}

\author{Jennifer~G.~Winters}
\affiliation{Center for Astrophysics ${\rm \mid}$ Harvard {\rm \&} Smithsonian, 60 Garden Street, Cambridge, MA 02138, USA}


\author[0000-0001-7416-7522]{David R.~Anderson}
\affiliation{Astrophysics Group, Keele University, Staffordshire, ST5 5BG, UK}
\affiliation{Centre for Exoplanets and Habitability, University of Warwick, Gibbet Hill Road, Coventry CV4 7AL, UK}

\author{C\'{e}sar Brice\~{n}o}
\affiliation{Cerro Tololo Inter-American Observatory, Casilla 603, La Serena, Chile} 

\author{Nicholas Law}
\affiliation{Department of Physics and Astronomy, The University of North Carolina at Chapel Hill, Chapel Hill, NC 27599-3255, USA}

\author[0000-0003-3654-1602]{Andrew W. Mann}
\affiliation{Department of Physics and Astronomy, The University of North Carolina at Chapel Hill, Chapel Hill, NC 27599-3255, USA}

\author{Xavier Bonfils}
\affiliation{Univ. Grenoble Alpes, CNRS, IPAG, 38000 Grenoble, France}

\author[0000-0002-8462-515X]{Nicola Astudillo-Defru}
\affiliation{Departamento de Astronom{\'i}a, Universidad de Concepci{\'o}n, Casilla 160-C, Concepci{\'o}n, Chile}

\author[0000-0002-4625-7333]{Eric~L.~N.~Jensen}
\affiliation{Dept.\ of Physics \& Astronomy, Swarthmore College, Swarthmore PA 19081, USA}

\author{Guillem Anglada-Escud{\'e}}
\affiliation{School of Physics and Astronomy, Queen Mary University of London, 327 Mile End Road, London, E1 4NS, UK}

\author[0000-0003-2058-6662]{George~R.~Ricker}
\affiliation{Department of Physics and Kavli Institute for Astrophysics and Space Research, Massachusetts Institute of Technology, Cambridge, MA 02139, USA}

\author[0000-0001-6763-6562]{Roland~Vanderspek}
\affiliation{Department of Physics and Kavli Institute for Astrophysics and Space Research, Massachusetts Institute of Technology, Cambridge, MA 02139, USA}

\author[0000-0001-9911-7388]{David~W.~Latham}
\affiliation{Harvard-Smithsonian Center for Astrophysics, 60 Garden St, Cambridge, MA 02138, USA}

\author[0000-0002-6892-6948]{Sara~Seager}
\affiliation{Department of Physics and Kavli Institute for Astrophysics and Space Research, Massachusetts Institute of Technology, Cambridge, MA 02139, USA}
\affiliation{Department of Earth, Atmospheric and Planetary Sciences, Massachusetts Institute of Technology, Cambridge, MA 02139, USA}
\affiliation{Department of Aeronautics and Astronautics, MIT, 77 Massachusetts Avenue, Cambridge, MA 02139, USA}

\author[0000-0002-4265-047X]{Joshua~N.~Winn}
\affiliation{Department of Astrophysical Sciences, Princeton University, 4 Ivy Lane, Princeton, NJ 08544, USA}

\author[0000-0002-4715-9460]{Jon~M.~Jenkins}
\affiliation{NASA Ames Research Center, Moffett Field, CA, 94035, USA}

\author{Gabor Furesz}
\affiliation{Department of Physics and Kavli Institute for Astrophysics and Space Research, Massachusetts Institute of Technology, Cambridge, MA 02139, USA}

\author[0000-0002-5169-9427]{Natalia~M.~Guerrero}
\affiliation{Department of Physics and Kavli Institute for Astrophysics and Space Research, Massachusetts Institute of Technology, Cambridge, MA 02139, USA}

\author[0000-0003-1309-2904]{Elisa Quintana}
\affiliation{NASA Goddard Space Flight Center, 8800 Greenbelt Road, Greenbelt, MD 20771, USA}

\author[0000-0002-6778-7552]{Joseph~D.~Twicken}
\affiliation{NASA Ames Research Center, Moffett Field, CA, 94035, USA}
\affiliation{SETI Institute, Mountain View, CA 94043, USA}

\author{Douglas~A.~Caldwell}
\affiliation{NASA Ames Research Center, Moffett Field, CA, 94035, USA}
\affiliation{SETI Institute, Mountain View, CA 94043, USA}

\author{Peter~Tenenbaum}
\affiliation{NASA Ames Research Center, Moffett Field, CA, 94035, USA}
\affiliation{SETI Institute, Mountain View, CA 94043, USA}

\author[0000-0003-0918-7484]{Chelsea~ X.~Huang}
\affiliation{Department of Physics and Kavli Institute for Astrophysics and Space Research, Massachusetts Institute of Technology, Cambridge, MA 02139, USA}
\affiliation{Juan Carlos Torres Fellow}

\author[0000-0002-4829-7101]{Pamela~Rowden}
\affiliation{School of Physical Sciences, The Open University, Milton Keynes MK7 6AA, UK}

\author[0000-0002-0149-1302]{B{\'a}rbara~Rojas-Ayala}
\affiliation{Departamento de Ciencias Fisicas, Universidad Andr{\'e}s Bello, Fernandez Concha 700, Las Condes, Santiago, Chile}

\correspondingauthor{Tianjun~Gan}
\email{gtj18@mails.tsinghua.edu.cn}

\begin{abstract}
    We report the first discovery of a thick-disk planet, \tar b (TOI-704b, TIC 260004324), detected in the \tess\ survey. \tar b transits a bright\ ($V$ = 12.19 mag, $K$ = 7.99 mag) and quiet M dwarf located $\rm 29.87\pm0.02\ pc$ away with a mass of $\rm 0.502\pm0.015\ M_{\odot}$ and a radius of $\rm 0.501\pm0.030\ R_{\odot}$. We validate the planet by combining space and ground-based photometry, spectroscopy, and imaging. The planet has a radius of $\rm 1.088\pm 0.064\ R_{\oplus}$ with a $\rm 3 \sigma$ mass upper-limit of $\rm 8.7\ M_{\oplus}$. We analyze the galactic kinematics and orbit of the host star \tar\ and find that it has a large probability\ ($\rm P_{thick}/P_{thin}=6482$) to be in the thick disk with a much higher expected maximal height\ ($\rm Z_{max}=1.8\ kpc$) above the Galactic plane compared with other \tess\ planet host stars. Future studies of the interior structure and atmospheric properties of planets in such systems using for example the upcoming {\it James Webb Space Telescope} (\jwst), can investigate the differences in formation efficiency and evolution for planetary systems between different Galactic components\ (thick and thin disks, and halo).
\end{abstract}


\section{Introduction}
Since \cite{Gilmore1983} first proposed subdivision between the thick disk and thin disk after studying the stellar luminosity function and Galactic stellar number density gradient, the study of the origin of Galactic disks has been a hot topic over the past few decades. Current theories postulate that the Milky Way\ (MW) is made up of several components: a thin disk, a thick disk, a halo and a bulge. Further studies indicate that solar neighbourhood stars are mostly members of the Galactic disk, with a small fraction belonging to the halo \citep{Buser1999,Juric2008,Bensby2014}. In general, compared with thin-disk stars, stars in the thick disk are older \citep{Bensby2005,Fuhrmann2008,Adibekyan2011}, have enhanced $\rm \alpha$-elements abundance and lower metallicity \citep{Prochaska2000,Reddy2006,Adibekyan2013} as well as hotter kinematic features \citep{Adibekyan2013,Bensby2014}, which could affect the planet formation efficiency \citep{Gonzalez1997,Neves2009}.
\par To date, more than 4000 exoplanets\footnote{\url{https://exoplanetarchive.ipac.caltech.edu/}} have been detected, thanks to successful surveys such as HATNet \citep{Bakos2004}, SuperWASP \citep{Pollacco2006}, and space-based missions including CoRoT \citep{Baglin2006}, {\it Kepler} \citep{Borucki2010} and K2 \citep{Howell2014}. However, few of the known exoplanets have been claimed to show thick-disk features \citep{Reid2007,FuhrmannHD155358,Neves2009,Bouchy2010,Campante2015}. The difference in planet formation and evolution between the thick and thin disks of the Milky Way is still a mystery. Interestingly, a recent work from \cite{Mctier2019} implies that planets in the solar neighborhood are just as likely to form around fast moving stars\ (thick-disk) as they are around slow moving stars\ (thin-disk). Because a common way to separate different components of the Milky Way relies on the spatial motion of stars, potential large biases may arise from radial velocity\ (RV) measurement limits as the RV survey of \gaia\ DR2 focuses on relatively bright stars\ ($G \lesssim 16.2$\ mag). Only $\sim$ 150 million stars have RV measurements \citep{Sartoretti2018} so kinematic information of most faint stars is still lacking.

\par The successful launch of the Transiting Exoplanet Survey Satellite (\tess, \citealt{Ricker2014}) opened a new era in this area, aiming at detecting small exoplanets around bright stars, and capable of discovering about $\sim 10^4$ planets during its primary mission \citep{Sullivan2015,Huang2018}. The \tess\ survey can provide a large sample of solar neighborhood transiting planets across the whole sky. All planet host stars are bright enough to have their RV measured by the \gaia\ survey. It will be an excellent opportunity to study the difference in the planet evolution between the thin and thick disks. 

\par Here we present the discovery of \tar b, an Earth-size planet on a short 3.1843-day orbit, transiting a nearby M1-type dwarf. It is the first planetary system detected in the Galactic thick disk during the two-year survey of \tess.

\par This paper is organized as follows: In Section \ref{obs}, we describe the space and ground-based observations. Section \ref{da} presents the analysis about the stellar characterization of \tar\ along with results of the joint fit. We focus on the tidal evolution in Section \ref{sec:tidelvol}. In Section \ref{tdc}, we discuss the thick-disk features of \tar. We conclude our findings in Section \ref{dc}.



\section{Observations}\label{obs}
\subsection{TESS}
\tar\ (TIC 260004324) falls in TESS's continuous viewing zone\ (CVZ) and it was observed with the two-minute cadence mode, spanning from 2018 July 25th to 2019 July 17th. Data ranges from Sector 1 to Sector 13 while excluding Sector 6, and it consists of a total of 229,712 exposures. 

\par Once images were transmitted to Earth, they were reduced by using the Science Processing Operations Center\ (SPOC) pipeline \citep{Jenkins2016} which was developed at NASA Ames Research Center based on Kepler mission's science pipeline. Transit planet search \citep[TPS;][]{Jenkins2002,Jenkins2017} was performed to look for transit signals and finally \tar\ was alerted on the MIT \tess\ Alerts portal\footnote{\url{https://tess.mit.edu/alerts/}} as a planet candidate, designated \tess\ object of interest (TOI) 704.01, with a period of 3.814 days, a transit depth of $\rm \sim 400\ ppm$, and a transit duration of $\rm \sim 1.4\ hrs$. 

\par We downloaded photometric data from the Mikulski Archive for Space Telescopes\ (MAST\footnote{\url{http://archive.stsci.edu/tess/}}) and used the 2-minute Presearch Data Conditioning Simple Aperture Photometry (PDCSAP) light curve from the SPOC pipeline for our transit analyses \citep{Stumpe2012,Smith2012,Stumpe2014}, which has been corrected for instrumental and systematic effects. To improve the precision of the light curve, we ignored data where the SPOC quality flag was non-zero. We performed the detrending by fitting a spline model to the raw light curve after masking out all transits\ (knots spaced every 0.5 days). We divided the light curve by the best-fit spline for normalization. 

\begin{figure*}[htbp]
\centering
\includegraphics[width=0.8\textwidth]{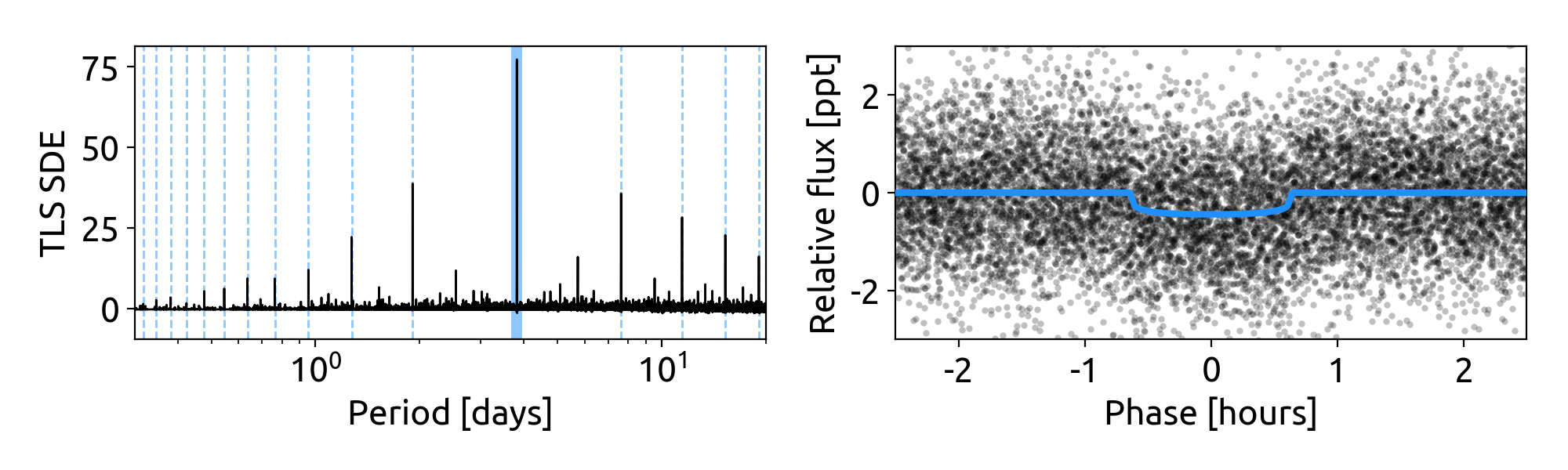}
\caption{Left: TLS power spectrum of the \tess\ photometry of \tar, with the detected orbital period indicated by a blue shaded region, as well as harmonics and sub-harmonics indicated by blue dotted lines. Right: The \tess\ photometry phase-folded on the detected orbital period, with the TLS transit model in blue; this model was subtracted from the data to search for additional transit signals but none were found.}
\label{tls}
\end{figure*}

To independently confirm the 3.814 day signal using all available \tess\ data (12 Sectors), we used the transit least-squares algorithm \citep[TLS;][]{Hippke2019} to search the light curve for transits. TLS uses a physically realistic model accounting for limb-darkening and nonzero ingress/egress duration, enabling it to detect shallower transits than BLS. We recovered the 3.814 day transits with a signal detection efficiency (SDE) of $\sim$75, and subtracted the TLS model from the data to search for additional planets (see Figure~\ref{tls}); several peaks with SDE moderately higher than 15 can be seen in the TLS power spectrum of the residuals, but they all appeared to be caused by noise. We concluded that no other significant transit signals exist in the \tess\ data besides the 3.814 day signal.

\begin{figure*}[htbp]
\centering
\includegraphics[width=17cm]{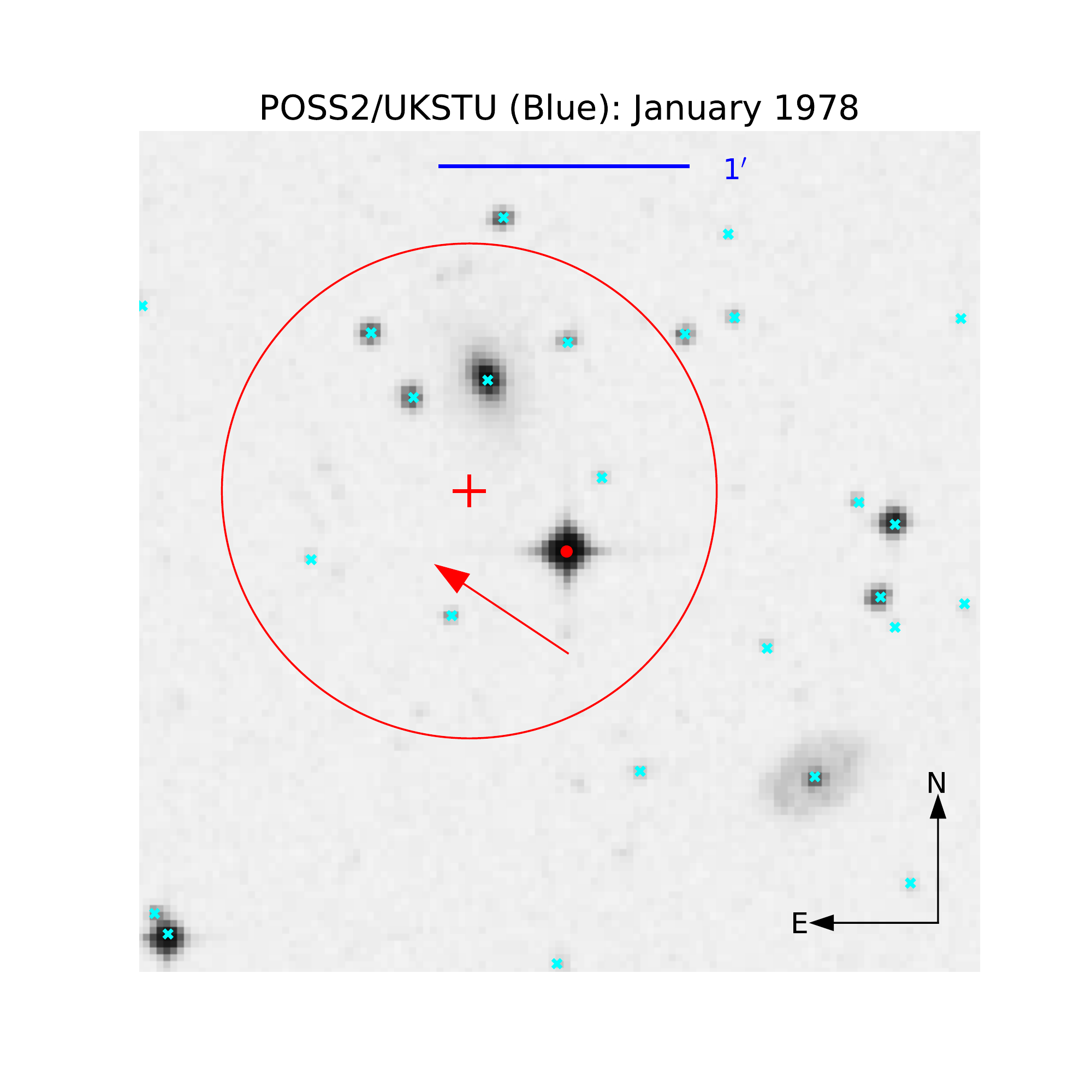}
\caption{The POSS2 blue image of \tar\ obtained in 1978. The red point is the location of \tar\ in POSS2 while the red cross represents its current position. The red circle indicates a region with a radius of 1$'$ around \tar. The red arrow indicates the direction of proper motion. Cyan points are stars within 2.5$'$ retrieved from \gaia\ DR2 that can potentially cause the \tess\ detection, all of which have been cleared by ground-based \lco\ photometry.}\label{FOV}
\label{image}
\end{figure*}

\begin{figure}[htbp]
\centering
\includegraphics[trim={0 0 0 0},clip,width=\columnwidth]{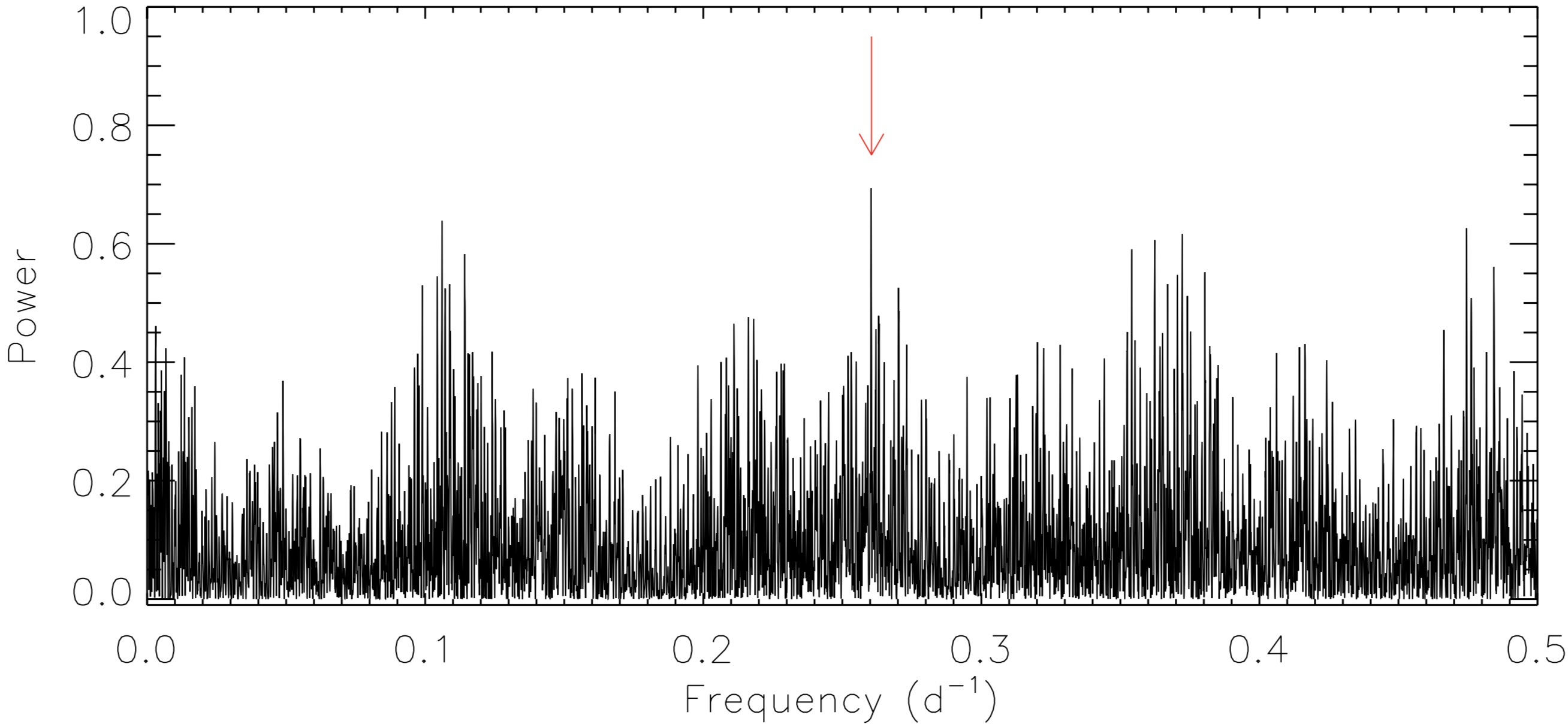}
\includegraphics[trim={0 0 0 0},clip,width=0.8\columnwidth]{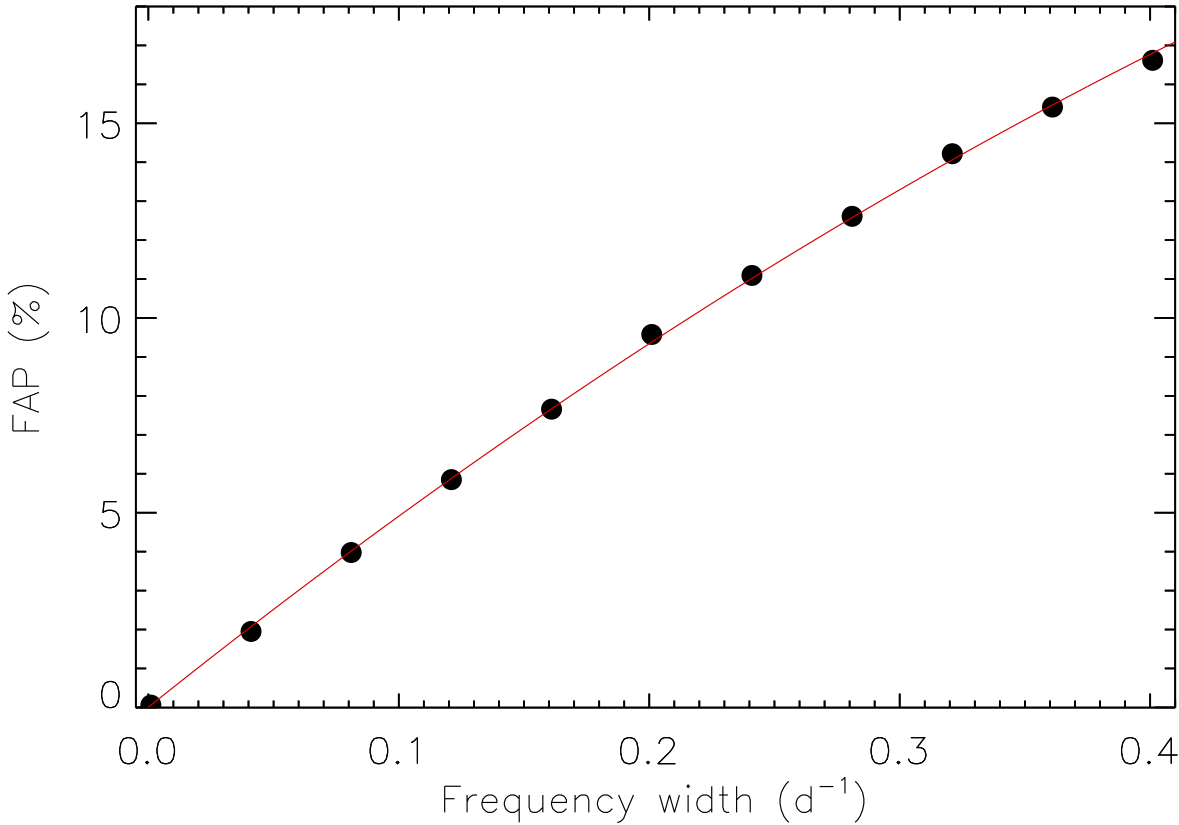}
\caption{Top: Generalized Lomb-Scargle periodogram of the HARPS RVs. The red arrow marks the orbital frequency of the transiting planet (f$_\mathrm{orb}$\,=\,0.262\,d$^{-1}$). Bottom: False-alarm probability computed in 11 different spectral ranges centered around the orbital frequency of the transiting planet (f$_\mathrm{orb}$\,=\,0.262\,d$^{-1}$) and with a width of 0.001, 0.041, 0.081, 0.121, 0.161, 0.201, 0.241, 0.281, 0.321, 0.361, and 0.401 d$^{-1}$. The red line marks the best fitting parabolic trend.}
\label{GLS}
\end{figure}


\subsection{Ground-Based Photometry}\label{gbp}
Though \tess\ has high photometric precision, due to its large pixel scale ($21''$ per pixel, \citealt{Ricker2014}), light from nearby stars is blended with the target. Nearby eclipsing binary\ (NEB) are a common source of false positives in \tess\ \citep{Brown2003,Sullivan2015} as they can cause transit-like signals on the target. Ground-based observations have two main goals: one is to reproduce the transit signal, the other is to look for nearby eclipsing binaries and check whether the signal is on the target \citep{Deeg2009}. 

In addition to \tess\ photometry, we also acquired two ground-based follow-up observations through 1m telescopes of the Las Cumbres Observatory Global Telescope Network\ (LCO)\footnote{\url{https://lco.global/}} \citep{Brown2013}, summarized in Table \ref{po}. We used the Sinistro cameras, which deliver a field of view (FOV) of $26' \times 26'$ with a plate scale of $\rm 0.389''$ per pixel. Data calibration was done by LCO's automatic BANZAI pipeline. Aperture photometry is performed by using AstroImageJ \citep{Collins2017}.

\par A full transit of \tar b was observed in the Sloan $r'$ band on 2019 August 24th at Siding Spring Observatory\ (SSO), Australia. The observation was obtained with 130s exposure time, aiming to rule out all potential faint nearby eclipsing binaries that may result in the \tess\ detection. We initially aimed at ruling out nearby EBs since the shallow transit depth (400 ppm) is challenging for ground telescopes to detect. Another similar egress observation in $r'$ but with 70s exposure time was done two orbital periods later at Cerro Tololo Inter-American Observatory\ (CTIO), Chile. In these observations we have examined all nearby stars within $2.5$ arcmin from the target with brightness difference down to $ \Delta T \sim 8.7 $ mag identified by Gaia\footnote{\url{https://www.astro.louisville.edu/gaia_to_aij/}} (See Figure \ref{FOV}). None of them showed variability (an eclipse) at an amplitude which could have led to the transit seen in TESS data when their light is blended with the target on TESS CCD.

\begin{table*}[htbp]
    \centering
    \caption{Summary of photometric observations for \tar}
    \begin{tabular}{clcccc}
        \hline\hline
        Facility       &Date &Exposure time(s) &Total exposures &Filter  & Summary\\\hline
        LCO 1m SSO Sinistro &2019 Aug 24 &130 &46 &$r'$ &full \\
        LCO 1m CTIO Sinistro &2019 Sep 1 &70 &92 &$r'$ &ingress\\
         \hline
    \end{tabular}
    \label{po}
\end{table*}


\subsection{High Resolution Spectroscopy}
Twenty-two spectra of \tar\ were collected with the High Accuracy Radial velocity Planet Searcher\ (HARPS, \citealt{Mayor2003}) on the ESO\,3.6\,m telescope at La Silla Observatory in Chile. The spectrograph has a resolving power of R\,$\approx$\,115,000 and covers the spectral range from 380\,nm to 690\,nm. These spectra were taken between UT 2003 December 15 to UT 2010 December 18 and are publicly available on the ESO Science Archive Facility\footnote{\url{http://archive.eso.org/wdb/wdb/adp/phase3_spectral/form}.}. We note that some of the RVs from those spectra were derived using the K5 template and the others with the M2 template.

\begin{table}[htbp]
     \centering
     \caption{HARPS RV measurements of \tar. Time-stamps are are given in barycentric Julian Date in the barycentric dynamical time.}
     \begin{tabular}{ccc}
         \hline\hline
         BJD$_\mathrm{TDB}$       &RV\ (m~s$^{-1}$) &$\sigma_{RV}$\ (m~s$^{-1}$) \\\hline
         2452988.75308 &1.30 &1.80 \\
         2452998.71510 &2.28 &2.42 \\
         2453007.72615 &0.38 &0.81 \\
         2453295.87376 &0.94 &1.95 \\
         2453834.51235 &-4.99 &1.43 \\
         2454430.82565 &-5.25 &1.91 \\
         2454431.76826 &0.84 &1.62 \\
         2454751.87135 &0.00 &5.35 \\
         2454803.72204 &4.98 &4.86 \\
         2454814.73847 &-4.50 &1.77 \\
         2454833.76771 &-6.42 &1.73 \\
         2454841.69264 &-2.28 &1.52 \\
         2454931.50924 &3.50 &1.53 \\
         2455218.75761 &-0.27 &1.44 \\
         2455538.64177 &5.24 &2.22 \\
         2455539.64365 &-3.78 &1.94 \\
         2455540.66332 &-3.39 &1.70 \\
         2455542.70589 &-0.79 &2.28 \\
         2455544.72265 &0.36 &1.76 \\
         2455546.63073 &3.07 &2.10 \\
         2455547.74183 &-0.79 &2.00 \\
         2455548.64140 &-1.84 &2.29 \\
          \hline
     \end{tabular}
     \label{harpsrv}
\end{table}

Here we used the Template Enhanced Radial velocity Re-analysis Application \citep[TERRA,][]{Anglada2012} software to homogeneously extract the Doppler measurements from the archival HARPS spectra. TERRA is considered to be more precise for M-dwarfs relative to the HARPS Data Reduction Software \citep[DRS;][]{Perger2017} whose results are on the HARPS archive. Table \ref{harpsrv} lists the HARPS-TERRA RVs and their uncertainties.
Time stamps are given in barycentric Julian Date in the barycentric dynamical time (BJD\,$_\mathrm{TDB}$).

We searched the HARPS-TERRA RVs for the Doppler reflex motion induced by the transiting planet. Figure~\ref{GLS} displays the generalized Lomb-Scargle periodogram \citep{Zechmeister2009} of the HARPS-TERRA RVs within the frequency range 0.0\,--\,0.5\,d$^{-1}$. The periodogram has its highest peak at the orbital frequency of the transiting planet (f$_\mathrm{orb}$\,=\,0.262\,d$^{-1}$). We assessed its false-alarm probability (FAP) following the bootstrap method described in \citet{Murdoch1993}. Briefly, we defined the FAP as the probability that the periodogram of fake data sets -- obtained by randomly shuffling the Doppler measurements, while keeping their time-stamps fixed -- has a peak higher than the peak observed in the periodogram of the HARPS-TERRA RVs. With a false alarm probability of FAP\,$\approx$\,30\,\%, the signal at f$_\mathrm{orb}$\,=\,0.262\,d$^{-1}$ is found not to be significant within the frequency range 0.0\,--\,0.5\,d$^{-1}$.

\par Yet, the \tess\ light curve provides prior knowledge of the possible presence of a Doppler signal at the transiting frequency. We therefore computed the FAP at the orbital frequency of the transiting planet, i.e., the probability that random data sets can produce a peak exactly at f$_\mathrm{orb}$\,=\,0.262\,d$^{-1}$ and whose power is higher than the power of the peak found in the periodogram of the HAPRS-TERRA RVs. To this aim, we first computed the FAP of 10$^5$ fake data sets in 11 different spectral ranges centered around f$_\mathrm{orb}$\,=\,0.262\,d$^{-1}$ and with arbitrary chosen widths\footnote{We note that the time resolution of the HARPS time-series -- defined as the inverse of the time baseline -- is 0.0004\,d$^{-1}$, which is 2.5 times lower then the smallest width used in our analysis.} of 0.001, 0.041, 0.081, 0.121, 0.161, 0.201, 0.241, 0.281, 0.321, 0.361, and 0.401 d$^{-1}$. We finally extrapolated the FAP in an infinitesimally narrow window centered around f$_\mathrm{orb}$\,=\,0.262\,d$^{-1}$ by fitting a quadratic trend to the 11 data points. We found a small false alarm probability of FAP\,=\,0.02\,\%, providing evidence for the existence of a significant Doppler signal at the transiting frequency of the planet.



\subsection{High Angular Resolution Imaging}
\par High-angular resolution imaging is needed to search for nearby sources that can contaminate the TESS photometry, resulting in an underestimated planetary radius, or be the source of astrophysical false positives, such as background eclipsing binaries.
\subsubsection{SOAR}
\begin{figure}[htbp]
\centering
\includegraphics[width=8.5cm]{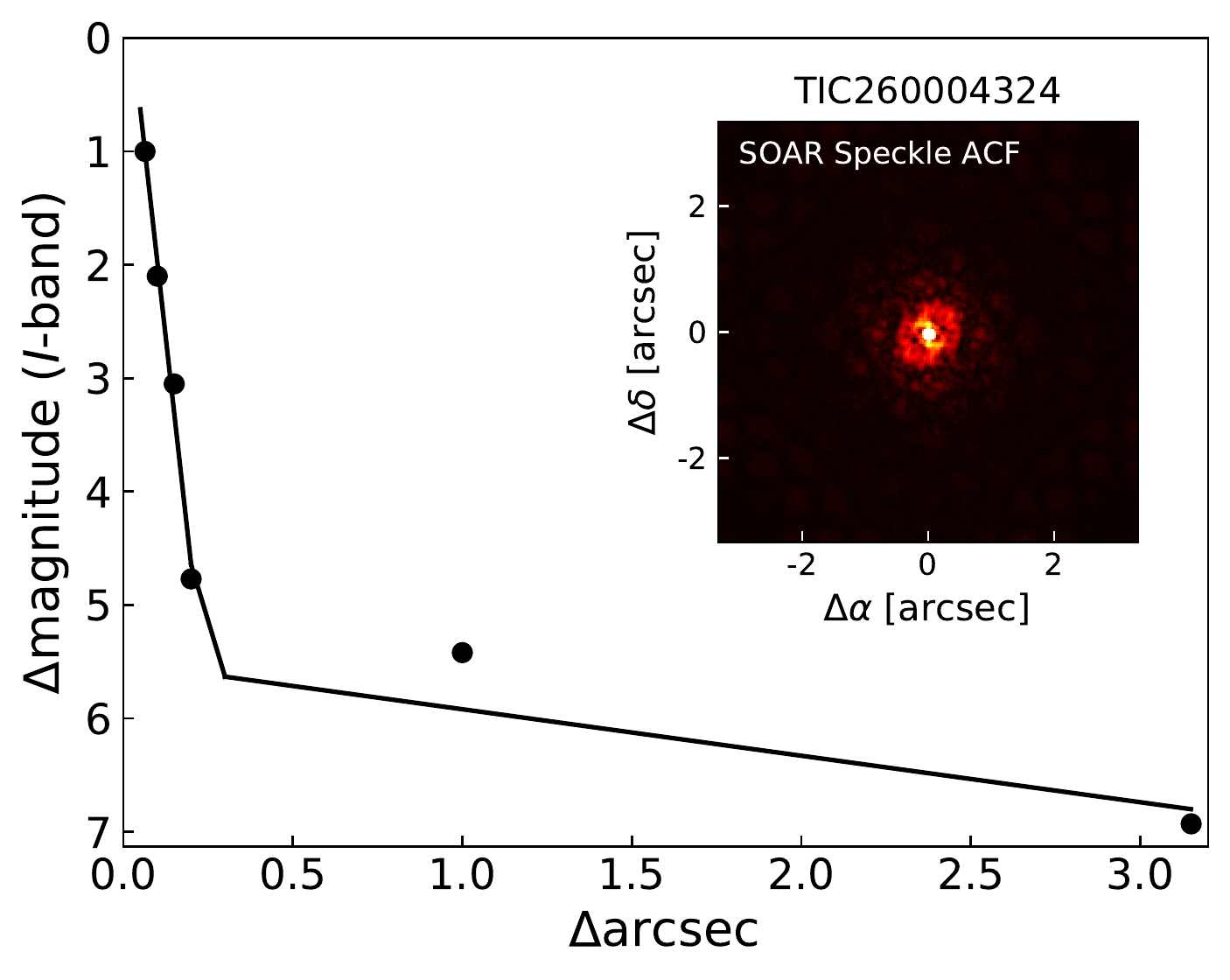}
\caption{Speckle auto-correlation function obtained in $I$-band using SOAR. The $5\sigma$ contrast curve for \tar\ is shown by the black points. The black solid line corresponds to the linear fit of the data, at separations smaller and larger than $\sim 0.2''$.} 
\label{SOAR}
\end{figure}

We searched for stellar companions to \tar\ with speckle imaging on the 4.1-m Southern Astrophysical Research\ (SOAR) telescope \citep{Tokovinin2018} on UT 16 October 2019, observing in a similar visible bandpass as TESS. The $5\sigma$ detection sensitivity and speckle auto-correlation function from the observation are shown in Figure \ref{SOAR}. No nearby stars were detected in the SOAR observations down to 5 magnitudes fainter than the target and as close as 0.2$\arcsec$ to \tar.

\subsubsection{Gemini-South}
\begin{figure}
\centering
\includegraphics[width=\columnwidth]{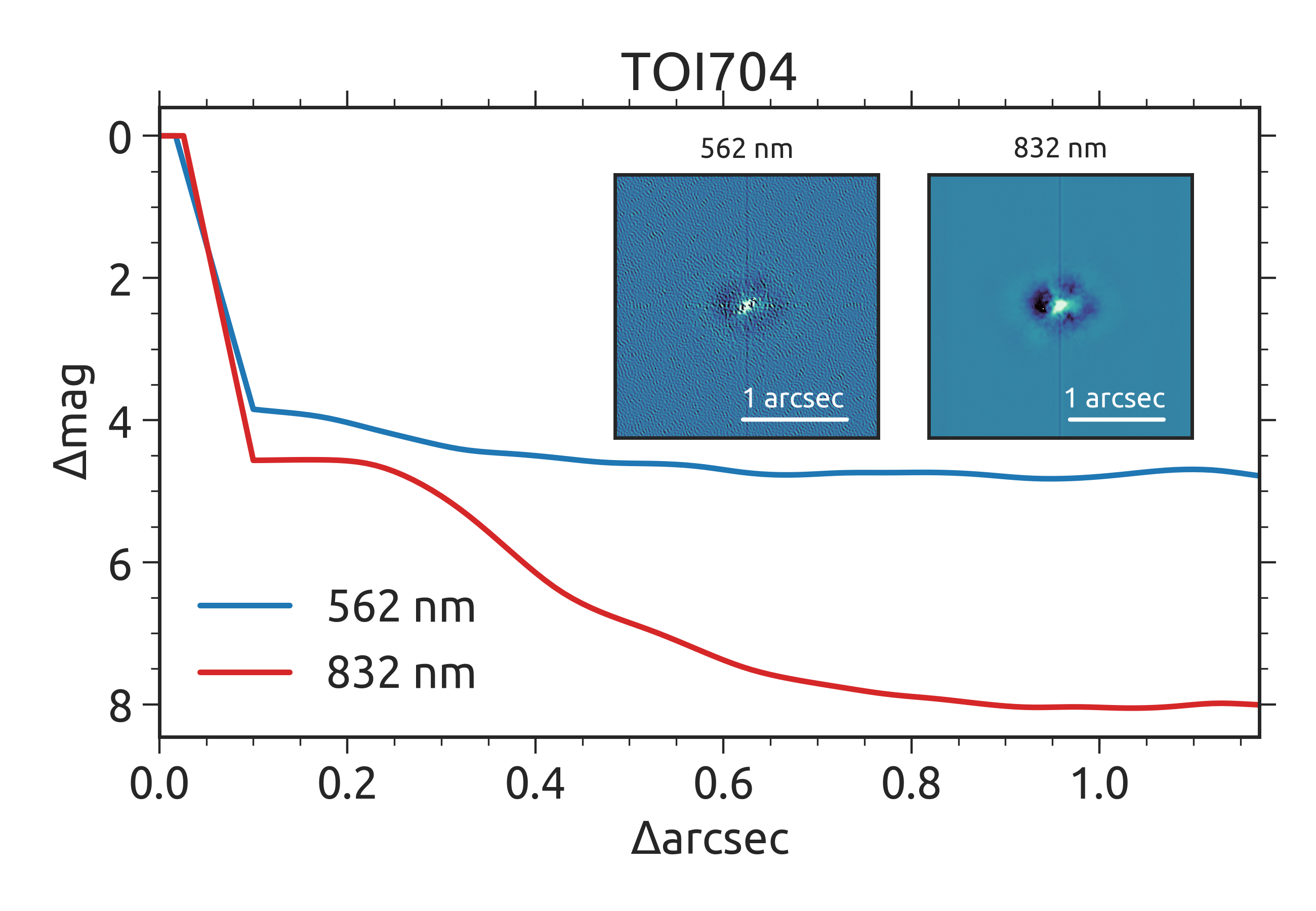}
\caption{Zorro speckle imaging and $5\sigma$ contrast curves of \tar\ at 562\,nm and 832\,nm. The data reveal that no companion star is detected from the diffraction limit (17\,mas) out to 1.75\,\arcsec within a $\Delta$\,m of 5 to 8.}
\label{zspeckle}
\end{figure}

\tar\ was also observed on UT 8 October 2019 using the Zorro speckle instrument on Gemini-South\footnote{\url{https://www.gemini.edu/sciops/instruments/alopeke-zorro/}}. Zorro provides simultaneously high-resolution speckle imaging in two bands, 562\,nm and 832\,nm, with output data products including a reconstructed image, and robust limits on companion detections \citep{Howell2011}. 
Figure~\ref{zspeckle} shows our results with corresponding reconstructed speckle images from which we find that \tar\ is a single star with no companions detected down to a magnitude difference of 5 to 8 mag from the diffraction limit\ (0.5 AU) to 1.75\,\arcsec\ (54 AU).

\section{analysis}\label{da}
\subsection{Stellar Characterization}
\subsubsection{Empirical Relation}\label{ER}
We used 2MASS $\rm m_{K_{S}}$ \citep{Cutri2003,skrutskie2006} and the parallax from \gaia\ DR2 \citep{GaiaHR2018} to calculate the $K_{S}$ band absolute magnitude $\rm M_{K_{S}}= 5.62 \pm 0.02$ mag. We estimated the bolometric correction to be $2.61\pm0.06$ mag through the empirical polynomial relation in \cite{Mann2015}. We obtained a bolometric magnitude $\rm M_{bol}=8.23\pm 0.06$ mag, leading to a luminosity of $\rm  L_{\star}=0.040\pm0.002\ L_{\odot}$. 
\par To compute the effective temperature $\rm T_{eff}$ of the host star, we applied two different methods. Following the polynomial relation between $\rm T_{eff} $ and $\rm V-J$ in \cite{Pecaut2013}, we obtained $\rm T_{eff}=3658\pm 103\ K$. We also determined $\rm T_{eff}$ based on the Stefan-Boltzmann law. First we estimated the radius of the host star $\rm 0.50\pm 0.03\ R_{\odot}$ using the $\rm R_{\star}$-$\rm M_{K_{S}}$ relation in \cite{Mann2015}. Then we derived $\rm T_{eff} = 3630\pm98\ K$, which agrees well with the result from the first method. 
\par We evaluated the mass of the host star $\rm M_{\star}=0.502\pm0.015\ M_{\odot}$ using Equation 2 in \cite{Mann2019} based on the $\rm M_{\star}$-$\rm M_{K_{S}}$ polynomial relation.

\subsubsection{Spectroscopic parameters}
Following \citet{Hirano2018}, we also used the co-added HARPS spectra (S/N\,=\,115 at 6000\,\AA) as input to \textbf{SpecMatch-Emp} \citep{Yee2017} to derive the stellar effective temperature $\rm T_{eff}$, radius $R_{\star}$, and iron abundance [Fe/H].
By matching the input spectrum to a high-resolution spectral library of 404 stars, this method yields $\rm T_{eff} = 3553 \pm 70 \rm\ K$, $\rm R_{\star} = 0.454 \pm 0.100\ R_{\odot}$, and $\rm [Fe/H] = -0.12 \pm 0.09$.

\subsubsection{SED Analysis}
As an independent check on the derived stellar parameters, we performed an analysis of the broadband spectral energy distribution\ (SED) together with the \gaia\ DR2 parallax in order to determine an empirical measurement of the stellar radius, following the procedures described in \cite{Stassun2016} and \cite{Stassun2017,Stassun2018}. We gathered the $U$, $B$, $V$ magnitudes from \cite{Mermilliod2006}, the $J$, $H$, $K_{\rm S}$ magnitudes from 2MASS Point Source Catalog \citep{Cutri2003,skrutskie2006}, four Wide-field Infrared Survey Explorer (WISE) magnitudes \citep{wright2010} and three \gaia\ magnitudes $\rm G$, $\rm G_{BP}$, $\rm G_{RP}$. Together, the available photometry spans the full stellar SED over the wavelength range 0.3--22~$\mu$m. 

\begin{figure}[htbp]
\centering
\includegraphics[width=8.5cm]{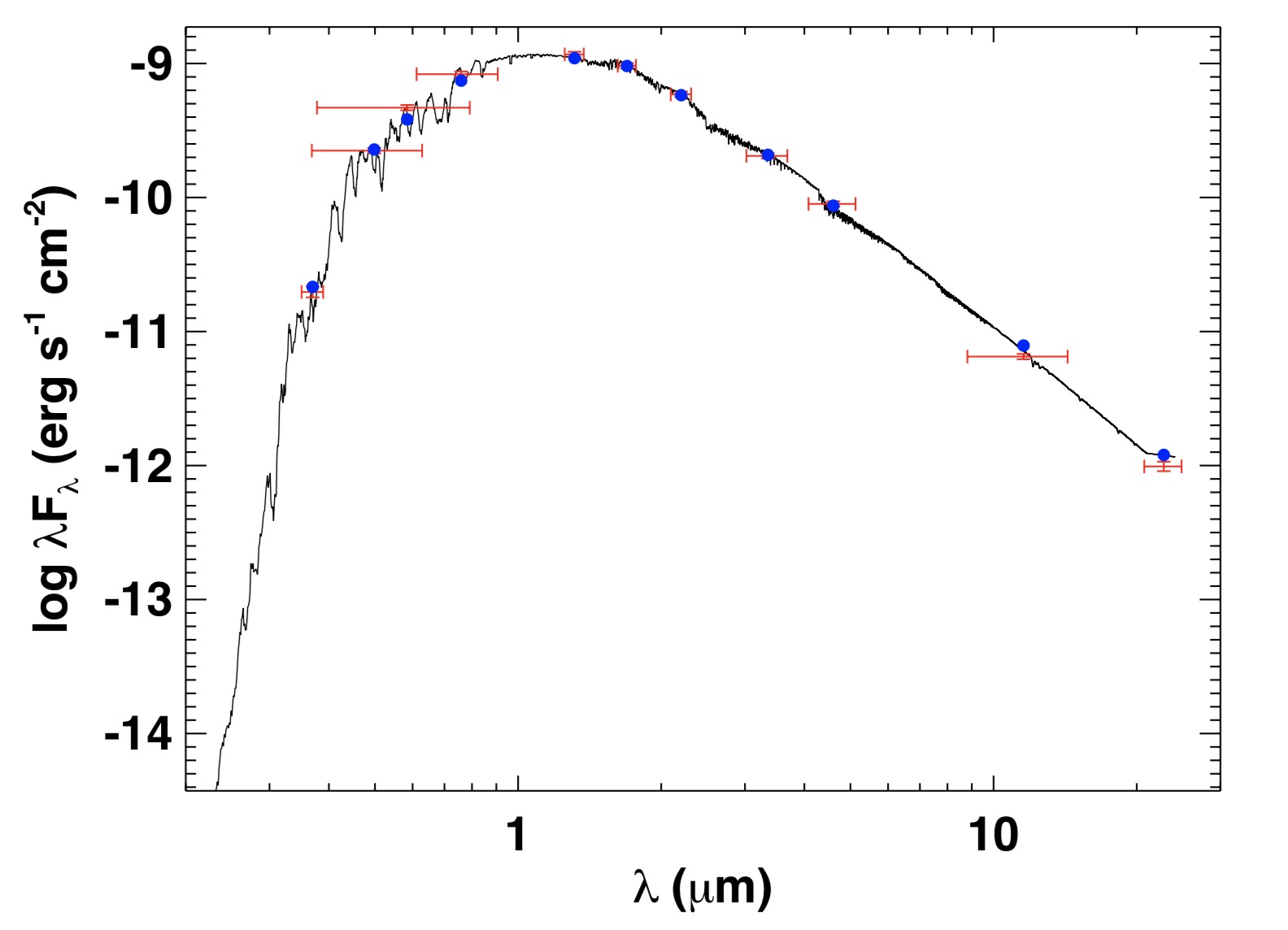}
\caption{The best SED fit for \tar. The Red symbols show the observed photometric measurements, where the horizontal bars represent the effective width of the passband. The Blue points are the predicted integrated fluxes at the corresponding bandpass. The black line represents the best-fit NextGen atmosphere model.}
\label{sed}
\end{figure}

We performed a fit using the NextGen stellar atmosphere models, with priors on effective temperature $\rm T_{eff}$ and metallicity\ ([Fe/H]) from the empirical relations and spectroscopy described above. We set the extinction\ $A_V$ to zero due to the proximity of the star. The best-fit SED is shown in Figure \ref{sed} with a reduced $\chi^2 = 2.5$, adopting $\rm T_{eff} = 3650 \pm 160$~K and [Fe/H] = $-0.12 \pm 0.09$. Integrating the model SED gives an observed bolometric flux of $\rm F_{bol} = 1.478 \pm 0.070 \times 10^{-9}$ erg~s$^{-1}$~cm$^{-2}$. Taking the $\rm F_{\rm bol}$ and $\rm T_{eff}$ together with the \gaia\ parallax, adjusted by $+0.08$~mas to account for the systematic offset reported by \citet{Stassun2018gaia}, we found a stellar radius of $\rm R = 0.502 \pm 0.044$~R$_\odot$ which is consistent with our result based on empirical relations in Section \ref{ER}. 

\par Combining all the results above, we adopted the mean values for effective temperature $\rm T_{eff}$ and stellar radius $\rm R_{\star}$. Together with the expected stellar mass, we found the mean stellar density $\rm \rho_\star = 5.6 \pm 2.7$~g~cm$^{-3}$. We list all stellar parameter values in Table \ref{starparam}. 

\begin{table}[htbp]\label{stellarinfor}
    \caption{Basic stellar parameters for \tar}
    \begin{tabular}{lll}
        \hline\hline
        Parameter       &Value       \\\hline
        \it{Star ID}                    \\
         2MASS                & J06042035-5518468 \\
         Gaia DR2            &5500061456275483776 \\
         TIC                     &260004324   \\
         TOI                     &704         \\
         LHS                     &1815        \\
         \it{Equatorial Coordinates} \\
         $\alpha\ (J2000)$                    &$\ \ $ 06:04:20.359 \\
         $\delta\ (J2000)$                    &$-$55:18:46.84    \\
         \it{Photometric properties}\\
         $\tess$\ (mag)           &$10.142\pm0.007$   &$\rm TIC\ V8^{[1]}$     \\
         $\gaia$\ (mag)           &$11.236\pm0.0007$   &Gaia DR2   \\
         \gaia\ BP\ (mag)           &$12.407\pm0.0017$   &Gaia DR2   \\
         \gaia\ RP\ (mag)           &$10.180\pm0.0014$   &Gaia DR2   \\
         $\rm B_{T}$\ (mag)                 &$14.027\pm0.502$   &Tycho-2  \\
         $\rm V_{T}$\ (mag)                 &$12.166\pm0.202$   &Tycho-2  \\
         $B$\ (mag)                    &$13.595\pm0.011$         &APASS \\
         $V$\ (mag)                    &$12.189\pm0.03$          &APASS \\
         $J$\ (mag)                    &$8.801\pm0.024$   &2MASS \\
         $H$\ (mag)                    &$8.209\pm0.047$   &2MASS \\
         $K_{S}$\ (mag)                &$7.993\pm0.020$    &2MASS \\
         WISE1 (mag)                   &$7.820\pm0.023$   &WISE \\
         WISE2 (mag)                   &$7.736\pm0.020$   &WISE \\
         WISE3 (mag)                   &$7.661\pm0.016$   &WISE \\
         WISE4 (mag)                   &$7.555\pm0.088$   &WISE \\
         \it{Astrometric properties}\\
         parallax (mas)              &$33.48\pm0.03$  &Gaia DR2  \\
         $\rm \mu_{\alpha}\ (mas~yr^{-1})$     &$681.73\pm0.05$   &Gaia DR2   \\
         $\rm \mu_{\delta}\ (mas~yr^{-1})$     &$342.13\pm0.06$   &Gaia DR2  \\
         RV\ (km~s$^{-1}$)                          &$42.22\pm0.25$ &Gaia DR2  \\
         \it{Derived parameters} \\
         Distance (pc)                &$29.87\pm0.02$  &This work     \\
         $\rm M_{\star}\ (M_{\odot})$ &$0.502\pm0.015$ &This work       \\
         $\rm R_{\star}\ (R_{\odot})$ &$0.501\pm0.030$ &This work       \\
         $\rm \rho_\star\ (g~cm^{-3})$ &$5.6\pm2.7$ &This work 
         \\
         $\rm log g_{\star}\ (cgs)$       &$4.77\pm0.03$  &This work        \\
         $\rm L_{\star}\ (L_{\odot})$ &$0.041\pm0.004$  &This work    \\
         $\rm T_{eff}\ (K)$           &$3643\pm142$  &This work       \\
         $\rm [Fe/H]$  &$-0.12\pm0.09$ &This work
         \\
         $\rm P_{rot}\ (d)$ &$47.8\pm0.7$ &This work 
         \\
         \hline\hline 
    \end{tabular}
    \begin{tablenotes}
    \item[1]  [1]\ \cite{Stassun2017tic,Stassun2019tic} 
    \end{tablenotes}
    \label{starparam}
\end{table}

\subsection{Joint Fit}\label{gm}
To simultaneously model the transits and radial velocity orbit, we used the EXOplanet traNsits and rAdIaL velocity fittER\ (EXONAIER, \citealt{Espinoza2016}). The transit model is created by {\textbf{batman}} \citep{Kreidberg2015} while the radial velocity orbit is modelled using {\textbf{radvel}} \citep{Fulton2018}. 
\par
\begin{figure*}[htbp]
\centering
\includegraphics[width=0.49\textwidth]{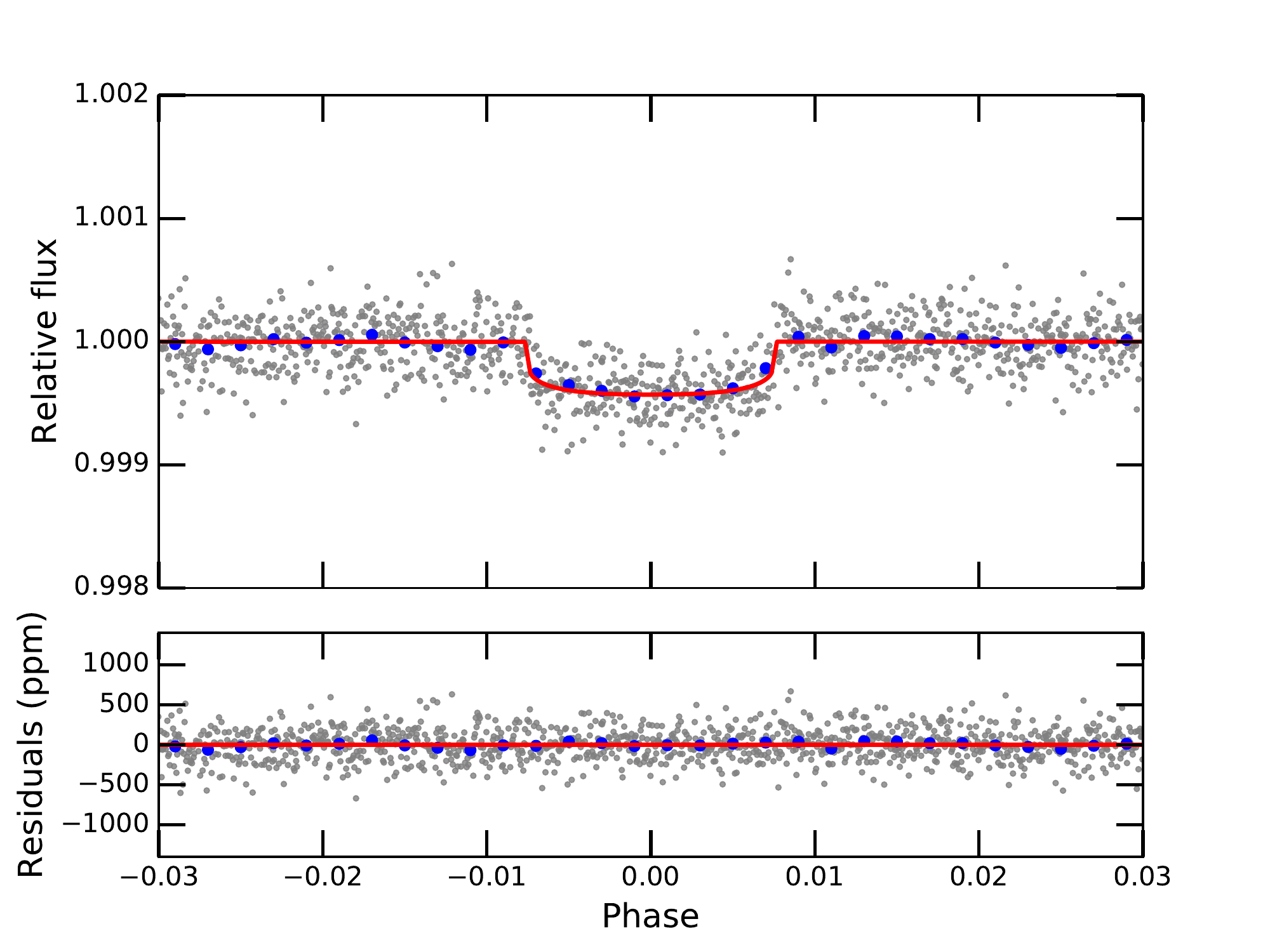}
\includegraphics[width=0.49\textwidth]{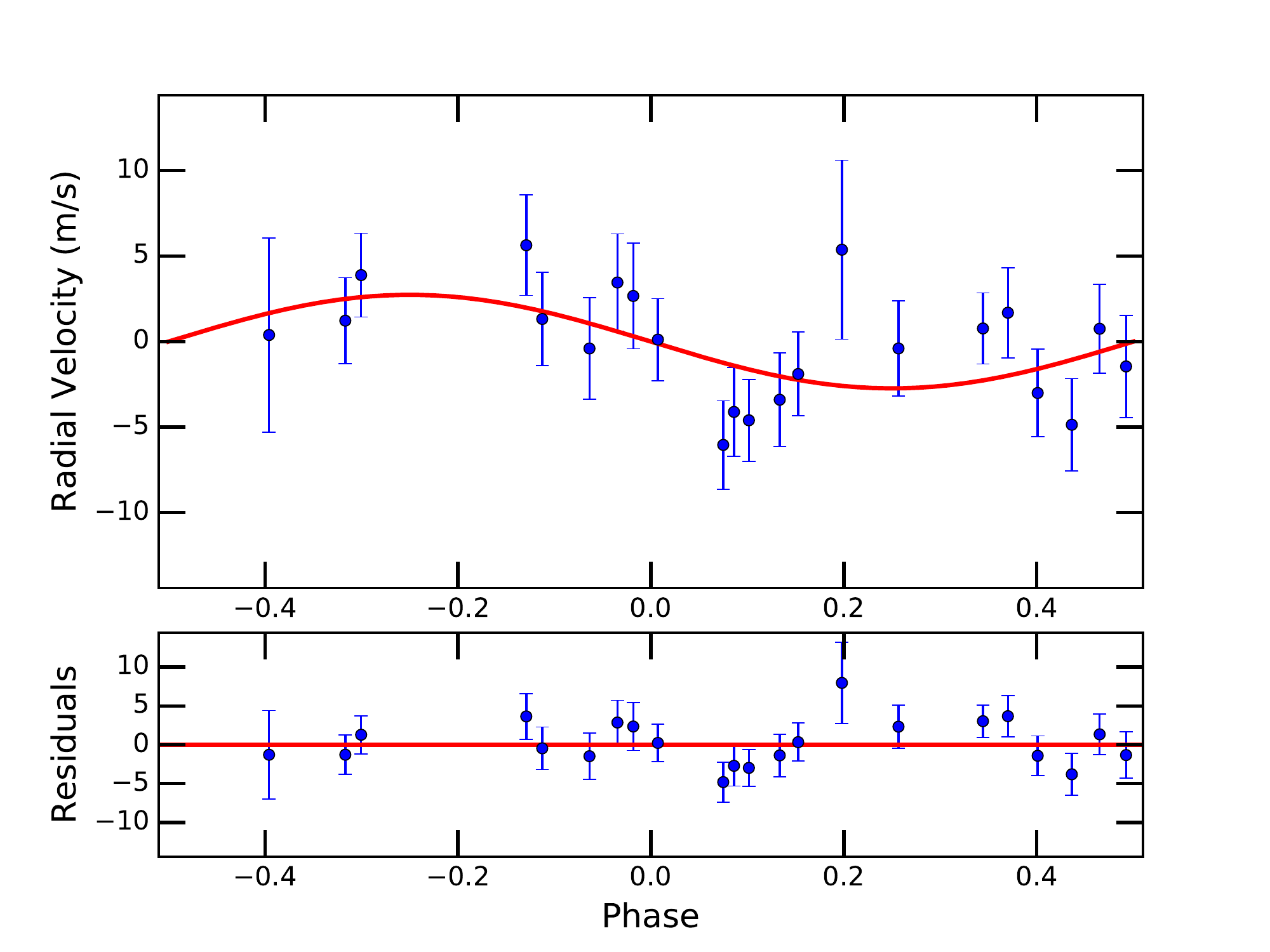}
\caption{Left: The phase-folded and normalized \tess\ photometric data. The binned light curves with different bin size are plotted with grey and blue points, respectively. The best-fit transit model is shown as a red solid line. Residuals are plotted below. Right: The phase-folded RV curve of \tar. Blue points represent RVs extracted from the HARPS spectra with the TERRA pipeline. The error bars are the quadrature sum of the instrument jitter term and the measurement uncertainties for all RVs. The best-fit model is shown as a red solid line. The residuals are shown below.}
\label{jointfit}
\end{figure*}

  Before we carried out the joint fit, we first created individual fit for \tess\ photometry-only and HARPS RV-only data sets with uniform priors, of which the posteriors are taken into consideration for further joint analysis. For the joint fit, we applied uniform priors for planet-to-star radius ratio\ ($\rm R_{P}/R_{\star}$), orbital inclination\ (i), two quadratic limb-darkening coefficients\ ($\rm q_{1}$ and $\rm q_{2}$) with an initial guess taken from \cite{Claret2018},  systemic velocity\ $\rm \gamma$, radial velocity semi-amplitude\ ({\it K}), and a normal prior for period\ (P), middle transit time\ ($\rm T_{0}$), and the separation between the host star and the planet in units of the stellar radius\ ($\rm a/R_{\star}$) based on the stellar radius and mass \citep{Sozzetti2007}. We applied the Markov Chain Monte Carlo\ (MCMC) analysis to determine the posterior probability distribution of the system parameters using the package {\textbf{emcee}} \citep{Foreman2013}. We first fitted a Keplerian orbit which gave an eccentricity of $0.4\pm0.2$, indicating the RV data set is insufficient to detect an eccentric orbit. Hence we assumed a circular orbit and fixed the orbital eccentricity to zero, which is expected given the short orbital period (see Section~\ref{sec:tidelvol}). The posterior of the semi-amplitude {\it K} is $\rm 2.7^{+0.9}_{-1.0}\ m~s^{-1}$, indicating that the companion of \tar\ has a mass $\rm 4.2\pm1.5\ M_{\oplus}$ with a $\rm 3\sigma$ upper-limit 8.7 $\rm M_{\oplus}$. The best-fit transit and RV models are shown in Figure~\ref{jointfit}. We list the resulting fitted parameters in Table \ref{planetparam} along with several derived physical parameters.

\subsection{Stellar rotation and activity}
\begin{figure}[htbp]
\includegraphics[width=0.48\textwidth]{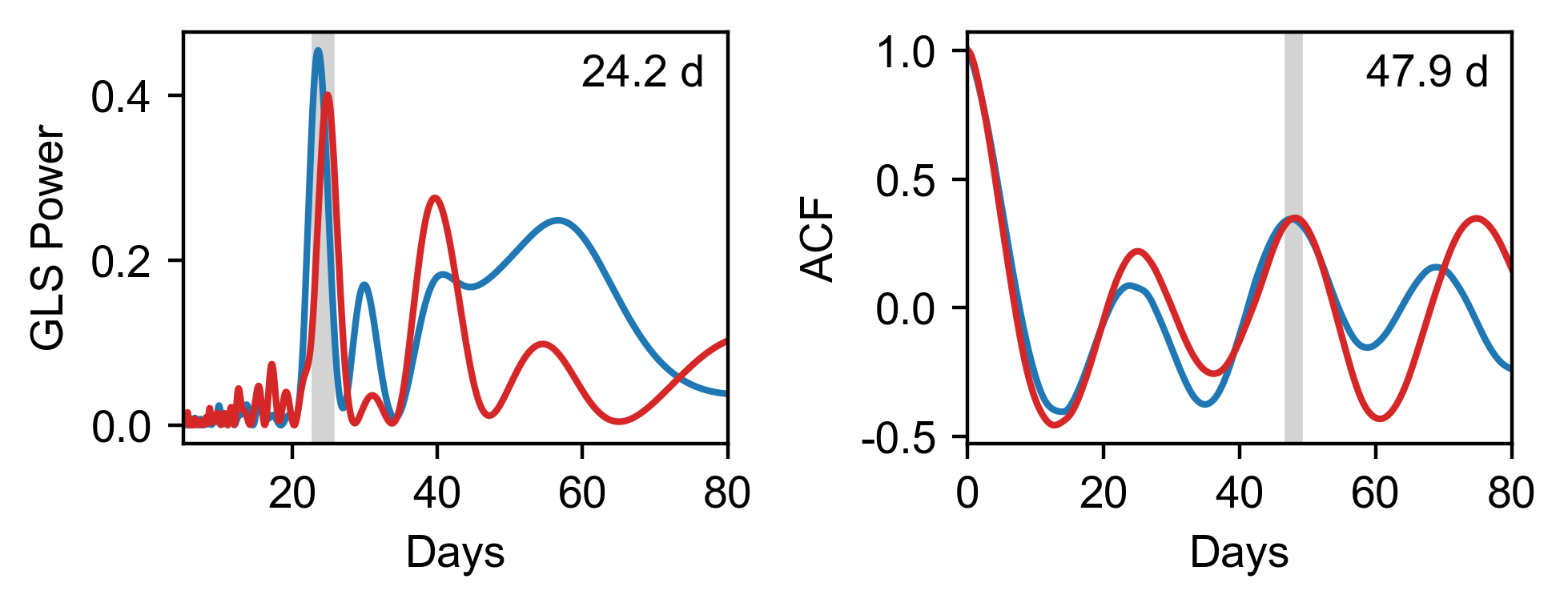}
\caption{GLS power spectrum (left) and auto-correlation function (right) of the PLD-corrected \tess\ photometry from Sectors 1-5 (blue) and 7-13 (red), with the peaks indicated by gray vertical shading. The mean values of the period from each data subset are annotated in the upper right.}
\label{rotation}
\end{figure}

\tess\ PDC SAP photometry is not always suitable for stellar variability studies, as the stellar variability can be removed by the PDC analysis. To search for rotational spot modulation in the \tess\ photometry, we used the {\bf lightkurve} package \citep{lightkurve} to produce systematics-corrected light curves from the \tess\ pixel data. {\bf lightkurve} implements a flavor of pixel-level decorrelation \citep[PLD;][]{Deming2015} to account for the correlated noise induced by the coupling of pointing jitter and intra-pixel gain inhomogeneities in the detector. We rejected outliers and normalized the PLD-corrected light curve from each \tess\ Sector to its median flux value, then further binned the data to a one day cadence for computational efficiency. We elected to analyze Sectors 1-5 and 7-13 independently because of the absence of data from Sector 6, which yielded two nearly-evenly sampled datasets. We computed the GLS periodogram \citep{Zechmeister2009} for each of the two data subsets, and found a clear peak in power at $\sim$24 days in both; less significant peaks can be seen at $\sim$40 and 55-60 days. Following \citet{Livingston2018}, we also computed the auto-correlation function (ACF) of each data subset, after linearly extrapolating the data to a uniformly spaced grid. For both data subsets the ACF exhibits a higher peak at $\sim$48 days, which suggests that the $\sim$24 day signal is the first harmonic of the rotation period (see Figure~\ref{rotation}); we concluded that the true stellar rotation period is $\sim$48 days. To estimate the uncertainty, we also modeled the full binned \tess\ time series as a Gaussian Process \citep{Rasmussen2005} with a quasi-periodic kernel, which enabled us to sample the posterior distribution via MCMC; we found the rotation period to be 47.8$\pm$0.7 days.

\begin{figure}[htbp]
\centering
\includegraphics[width=8.5cm]{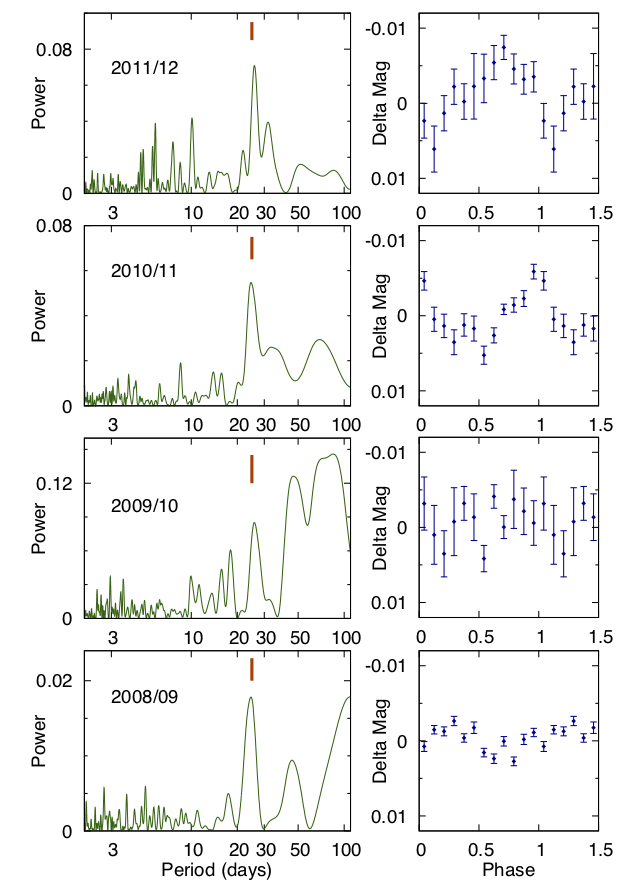}
\caption{Periodograms of the WASP-South data in each observing season, along with (right) folds of the data on the 24.9-d modulation (marked by orange ticks).} 
\label{wasp}
\end{figure}

\tar\ was also observed by WASP-South over the period of 2008 to 2012 for a typical duration of 150 days in each year. WASP-South is an array of 8 cameras combining 200-mm f/1.8 lenses with 2k$\times$2k CCDs and observing with a broad-band filter giving a 400-700 nm bandpass \citep{Pollacco2006}. Each visible field was monitored with a cadence of $\sim$\,15 mins on every clear night, accumulating 50,000 data points on \tar. The light curves from each observing season were searched for rotational modulations using the methods described in \cite{Maxted2011}. For \tar, we found a persistent modulation with a period of 24.9 $\pm$ 1.1 d with an amplitude of 2 to 8 mmag\ (Figure \ref{wasp}) and a false-alarm probability below 1\%. This is consistent with the signal found in the \tess\ data, and confirms that the signal is likely caused by rotation as it is persistent for multiple years. Future \tess\ data to be obtained during the \tess\ Extended Mission will allow better identification to the correct rotation period of this target. 

\begin{figure}[htbp]
\centering
\includegraphics[width=0.6\columnwidth]{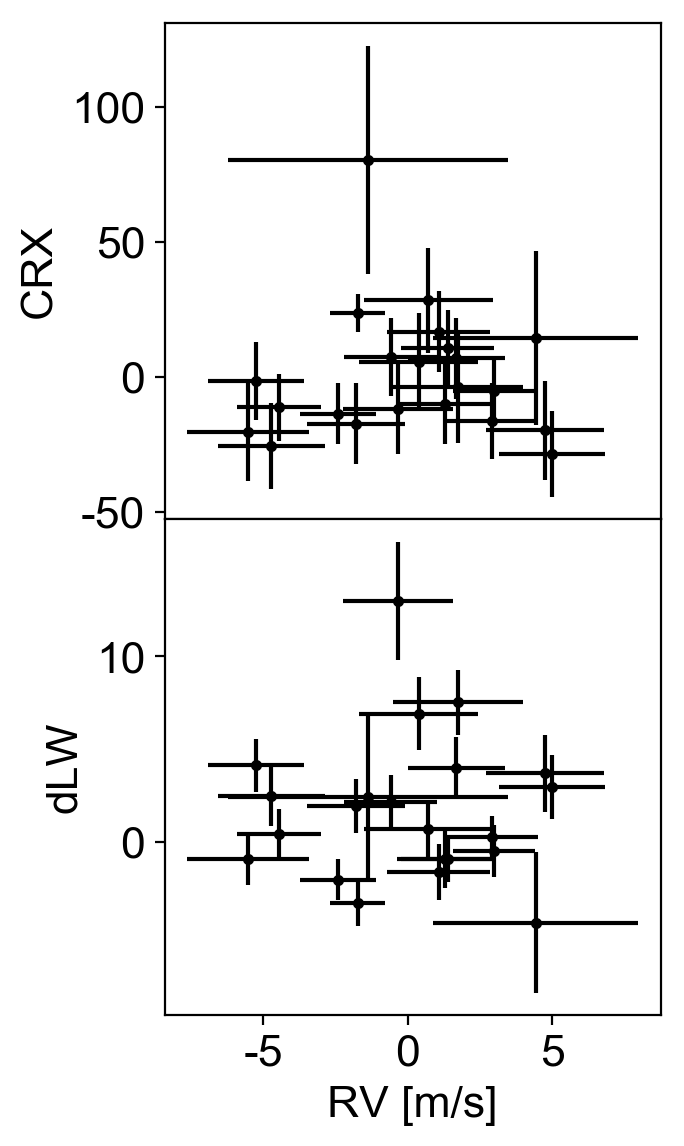}
\caption{Chromatic index (CRX) and differential line width (dLW) as a function of RV extracted from the HARPS spectra by the SERVAL pipeline.}
\label{serval}
\end{figure}

To assess stellar activity levels spectroscopically, we also extracted the chromatic index (CRX) and differential line width (dLW) indicators from the HARPS spectra using the publicly available SpEctrum Radial Velocity AnaLyser pipeline \citep[SERVAL;][]{Zechmeister2018}. CRX summarizes the wavelength dependence of the RVs, and dLW is an alternative to the commonly used FWHM. The apparent lack of a significant correlation between the activity indicators and the RVs suggests that the RVs are not dominated by stellar activity (see Figure~\ref{serval}). The observed RV scatter is therefore likely caused primarily by the Doppler signal induced by the planet, consistent with the detection of a peak in the GLS periodogram at the frequency of the orbital period (Figure~\ref{GLS}).

\subsection{False Positive Analysis}
As we previously discussed in Sec.~\ref{gbp}, there are several scenarios which can cause a false positive --- a transit-like signal in the \tess\ data that does not originate from a transiting star-planet system. We considered all data we have obtained and carefully ruled out the false positive scenarios below following \cite{Vanderspek2019}, \cite{Crossfield2019}, and \cite{Shporer2019}. 
\\
\par 1.\ {\it Detection is caused by instrumental artifact}: 
\par We excluded this possibility because periodic transit signals were found in all 12 \tess\ sectors in which this target was observed, and in each sector the target was located at different CCD position.
\\
\par 2.\ {\it \tar\ is a stellar eclipsing binary}:
\par Our HARPS RV data did not show a significant RV variability at the few m~s$^{-1}$ level. The $\rm 3\sigma$ mass upper-limit has also ruled out this scenario\ (Section \ref{gm}).
\\
\par 3.\ {\it Light from a nearby eclipsing binary is blended with \tar}:
\par Our two ground-based observations from \lco\ have cleared all nearby \gaia\ stars\ ($\rm \Delta T \sim 8.7$ mag) within $\rm 2.5'$ through the NEB analysis\ (Section \ref{gbp}). We did not find any obvious variation of those stars which indicates this cannot be the case. We have also made sure that the scatter in light curves of nearby stars is smaller than the expected eclipse depth given the brightness difference between the nearby star and the target.
\\
\par 4.\ {\it Light from an unassociated distant eclipsing binary or a transiting planet system fully blended with \tar} :
\par Thanks to the high proper motion of \tar\ ($\rm \sim 760\ mas~yr^{-1}$), we can easily reject this scenario by checking images from other surveys decades ago. We did not see any other stars that are bright enough to cause the transit seen in \tess\ data at the current position of \tar, as shown in Figure \ref{image}, thus this possibility is excluded. 
\\
\par 5.\ {\it \tar\ has a stellar binary companion on a wide orbit and that binary companion is the origin of the transit signal}:
\par Photometric data from multiple sectors of \tess\ offered us an opportunity to deliver precise duration of transit ingress/egress and the time from first-to-third contact during the transit event. Assuming a symmetric light curve, we have 
\begin{equation}
    \frac{\rm \tau_{12}}{\rm \tau_{13}}=\frac{\rm t_{T}-t_{F}}{\rm t_{T}+t_{F}},
\end{equation}
where $\rm t_{T}$ and $\rm t_{F}$ are the total and in-transit duration (2nd to 3rd contacts), respectively. \cite{Seager2003} gave the upper-limit of the radius ratio of the transiting planet:
\begin{equation}
    \frac{\rm R_{p\_real}}{\rm R_{\star}} \leq \frac{\rm \tau_{12}}{\rm \tau_{13}}.
\end{equation}
We constrained the relative flux drop if the signal is from an unresolved star:
\begin{equation}
    \frac{\rm \Delta F}{f_{b}}=\left(\frac{\rm R_{p\_real}}{\rm R_{\star}} \right)^2 \leq \left(\frac{\rm \tau_{12}}{\rm \tau_{13}} \right)^2 = 0.06\%.
\end{equation}
Given the $\rm 3\sigma$ lower-limit on the exact transit depth from global modelling, the blended star has to contribute at least 50\% of the total flux in the TESS aperture: 
\begin{equation}
    \frac{\rm \Delta F}{f_{s}+f_{b}} \geq 0.03\%;
\end{equation}
\begin{equation}
    \frac{f_{b}}{f_{s}+f_{b}} \geq 50\%,
\end{equation}
where $f_{s}$ and $f_{b}$ are the source flux and blending flux.
We excluded this scenario mainly based on the following reasons: 
\\
(1) According to this scenario the blending star is expected to have $\rm >50\%$ contribution to the \tess\ flux, but, \gaia\ and high resolution images show a non-detection of a nearby star at a few arcsec from the target. \\
(2) A star that is comparable in brightness to the target would make the spectrum appear double-lined but we do not see this phenomenon in the spectrum from HARPS. \\
(3) A star that is comparable in brightness to the target would cause the target to appear brighter for its distance. Since the distance is given by the \gaia\ DR2 parallax and $\rm T_{eff}$ is constrained by the SED, a blended star with comparable brightness will make the target appear too bright given its distance for a main sequence star, which is not the case. 


\section{Constraints from tidal evolution}\label{sec:tidelvol}


We estimated the timescales for circularization and tidal decay using the equilibrium tide model from \citet{hut81}. We integrated the secularly averaged equations for the eccentricity and semimajor axis of the planet (namely, equations 9 and 10 from \citealt{hut81}) using the midpoint method. We neglected the evolution of the planetary spin, since the spin angular momentum of the planet is too small with respect to the orbital angular momentum to affect the orbit significantly. Given the upper limit estimate for the mass $\mathrm{M}_{\mathrm{P}}$ of LHS 1815b and the intrinsic uncertainty of tidal efficiency parameters (the time-lag $\tau$ or the tidal quality factor $Q'$), we have explored different tidal evolution models in the range $10\,\mathrm{s} < \tau < 1000 \,\mathrm{s}$ and $ 1\,\mathrm{M_{\oplus}} < \mathrm{M}_\mathrm{P} < 6 \,\mathrm{M_{\oplus}}$. 
This range of time-lag is appropriate for planets with rocky composition \citep{soc2012}. For low tidal efficiency ($\tau = 10\,\mathrm{s}$), circularization takes longer than 10 Gyr regardless of the mass of the planet, while for high tidal efficiency ($\tau = 1000\,\mathrm{s}$) the planetary orbit is always completely circularized within 10 Gyr, with small planetary masses ($\mathrm{M}_\mathrm{P} < 3  \,\mathrm{M_{\oplus}}$) circularizing within 1 Gyr. On the other hand, at moderate tidal efficiency ($\tau = 100 \,\mathrm{s}$) the circularization timescale is sensible to the planetary mass. For $\tau = 100\,\mathrm{s}$ and $\mathrm{M}_\mathrm{P} < 3 \,\mathrm{M_{\oplus}}$ the planetary orbit reaches $e \lesssim 0.05$ within 10 Gyr, while it retains some eccentricity for higher planetary masses ($\mathrm{M}_\mathrm{P} > 3 \,\mathrm{M_{\oplus}}$).

\begin{figure}
    \includegraphics[trim={0 0 0 0},clip,width=0.5\textwidth]{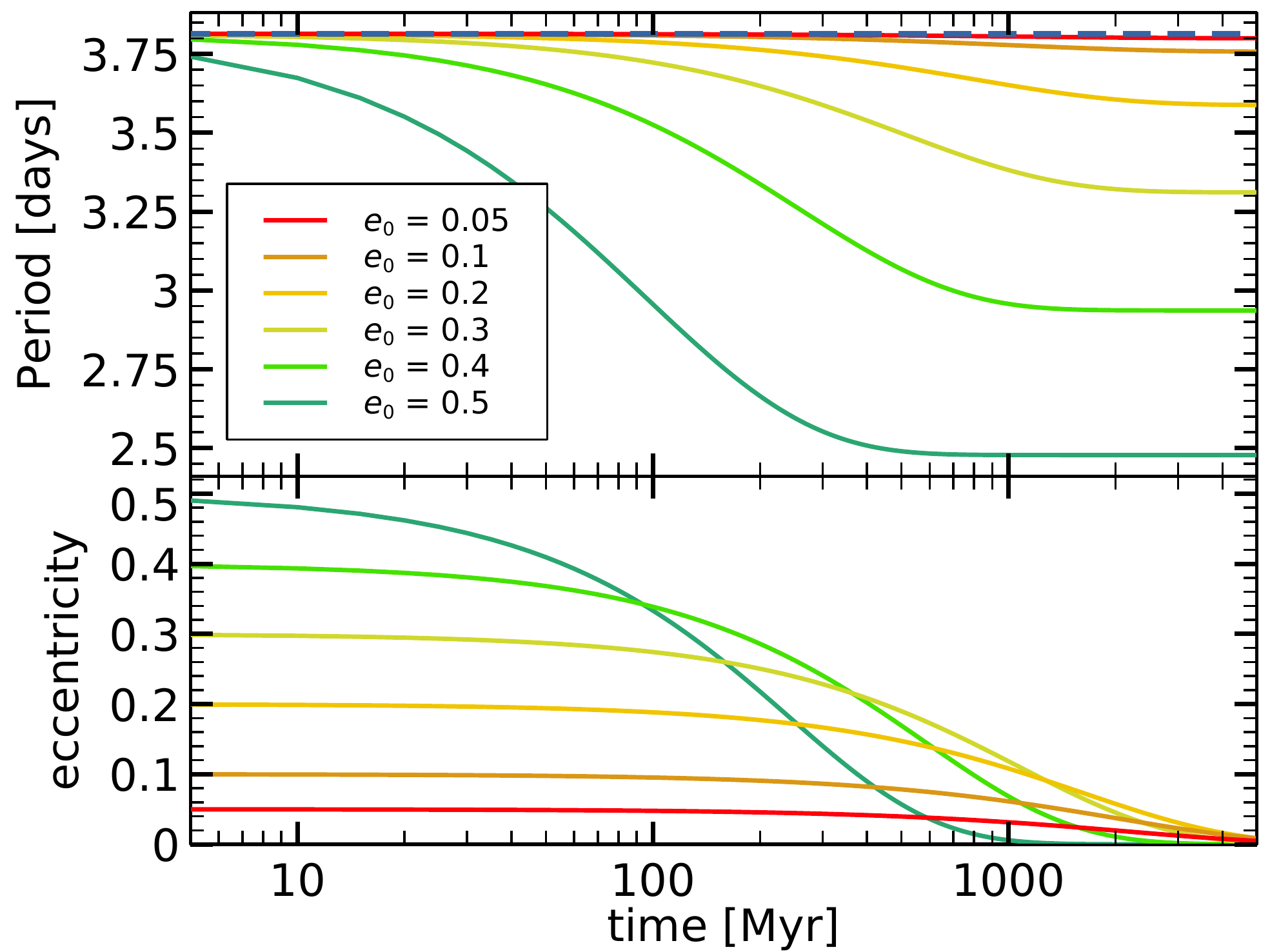}
    \includegraphics[trim={0 0 0 0},clip,width=0.5\textwidth]{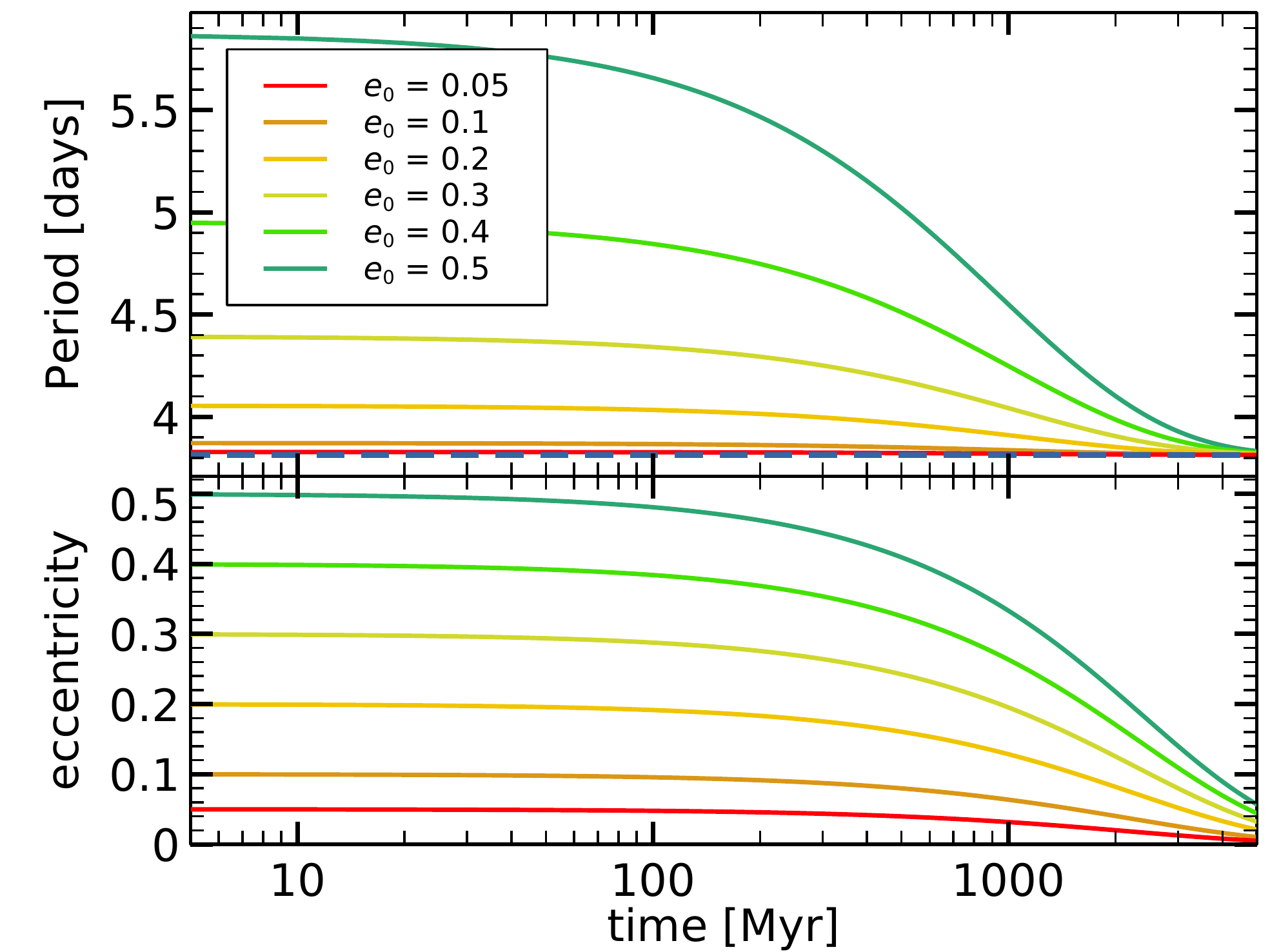}
    \caption{Period and eccentricity as a function of time for different initial eccentricities. The blue dashed line indicates the 5$\sigma$ error on the inferred period, which is smaller than the thickness of the line. Top panel: starting period equal to the observed one. Bottom panel: the initial period chosen so that the final period after circularization matches the observed one.}
    \label{fig:tidevol}
\end{figure}

\autoref{fig:tidevol} shows the evolution of orbital period and eccentricity of the planet, assuming $\mathrm{M}_\mathrm{P} = 4.5 \,\mathrm{M_{\oplus}}$, a constant time-lag of $\tau = 300\,\mathrm{s}$, and an apsidal constant of $k_{A}=0.3$, corresponding to a tidal quality factor of $Q' = 5 \times 10^2$.  In the top panel of \autoref{fig:tidevol} the planet has an initial period equal to the currently observed one, and different initial eccentricities. Initial eccentricities lower than 0.5 will be dissipated within about 5 Gyr, the lower the eccentricity, the longer the circularization time. However, as the eccentricity is dissipated, the orbital period decays so that the final period does not match the observed one. Specifically, the orbital period would mismatch the observed one within 100--200 Myr for all eccentricities $e \gtrsim 0.05$.

In the bottom panel of \autoref{fig:tidevol} we show the evolution for different initial periods so that the final period after circularization matches the present one. By 5 Gyr all the periods have reached the final value of $3.81433$ days with $e \lesssim 0.05$. If the system was younger than 5 Gyr, it would not have time to circularize unless the initial eccentricity was $e \lesssim 0.1$. Alternatively, it might be argued that the planet has not circularized yet. However, as shown by the top panel of \autoref{fig:tidevol}, any residual eccentricity higher than $0.05$ at $3.81433$ days would make the planet decay within $100$ Myr.

Ultimately, constraints on the age of the system would help narrowing down the possible range of eccentricities of the planet. If the system is very young ($\rm {\sim}100\ Myr$), the eccentricity is largely uncertain since the planet must be currently undergoing tidal circularization. Conversely, if the system is old ($>$5 Gyr), tidal circularization is mostly over and the eccentricity at present day is likely less than $0.05$. Note also that the eccentricity could be excited by another undetected planet, a possibility that we have neglected in our analysis.

\section{Thick-Disk Characteristics}\label{tdc}
\par We confirmed the thick-disk nature of \tar\ mainly on the basis of its kinematic information. In general, thick-disk stars are kinematically hotter\ (larger velocity dispersions) than stars that belong to the thin disk. We converted radial velocities and proper motions from \gaia\ DR2 to 3D velocities U, V and W\footnote{U, V, W are positive in the directions of Galactic center, Galactic rotation and the North Galactic Pole.} using the distance of d = 29.87$\pm$0.02 pc from our SED fit based on the method described in \cite{Johnson1987}. To relate the space velocities to the Local Standard of Rest (LSR), we adopted solar velocity components relative to the LSR\ ($\rm U_{\odot}$, $\rm V_{\odot}$, $\rm W_{\odot}$) = ($9.58$, $10.52$, $7.01$) km~s$^{-1}$ obtained by LAMOST \citep{Tian2015}. We determined the three-dimensional Galactic space motion of ($\rm U_{LSR}$, $\rm V_{LSR}$, $\rm W_{LSR}$) = ($-34.34\pm0.04$, $-71.47\pm0.22$, $76.26\pm0.14$) km~s$^{-1}$. 
\par
\begin{figure}[htbp]
\centering
\includegraphics[width=9cm]{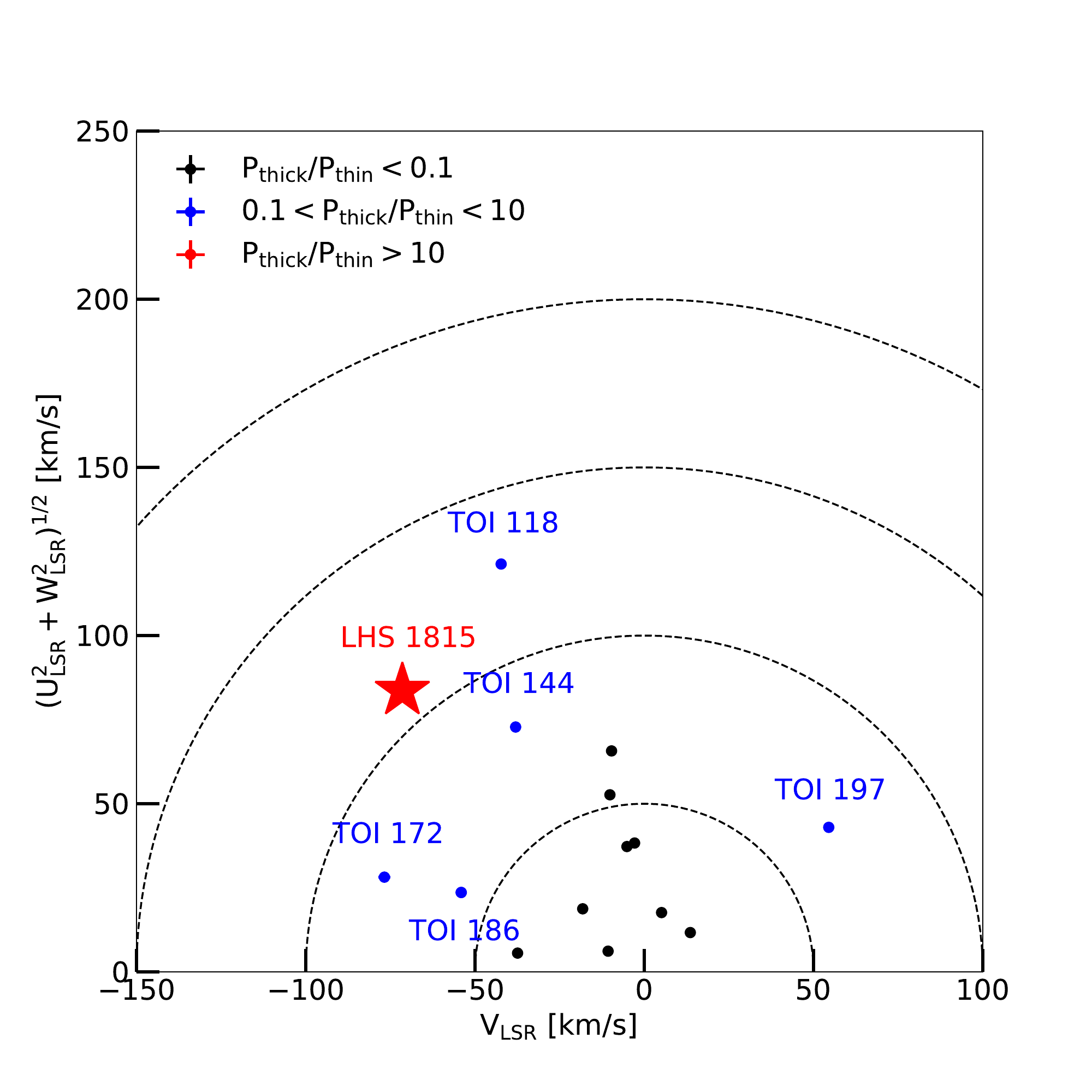}
\caption{The Toomre plot for all \tess\ host stars with planets. Different color represent different ranges of relative probability. Our target is shown as a red star. }
\label{toomre}
\end{figure}

 To judge which stellar component \tar\ belongs to, we employed the kinematical criteria first mentioned in \cite{Bensby2003} by assuming the Galactic space velocities $\rm U_{LSR}$, $\rm V_{LSR}$, and $\rm W_{LSR}$ of the stellar populations have Gaussian distributions:
\begin{equation}
    f=\rm k\times exp(-\frac{\rm {U_{LSR}}^{2}}{2{\sigma_{U}}^{2}}-\frac{\rm {(V_{LSR}-V_{asym})}^{2}}{2{\sigma_{V}}^{2}}-\frac{\rm {W_{LSR}}^{2}}{2{\sigma_{W}}^{2}}),
\end{equation}
\\where 
\begin{equation}
    \rm k=\frac{1}{(2\pi)^{3/2}\sigma_{U}\sigma_{V}\sigma_{W}}
\end{equation}
\\is a normalization constant, $\rm \sigma_{U}$, $\rm \sigma_{V}$ and $\rm \sigma_{W}$ represent velocity dispersion for 3D velocity components while $\rm V_{asym}$ is the asymmetric drift. We applied related parameters from \cite{Bensby2014} for solar-neighborhood stars and calculated relative probability $\rm P_{thick}/P_{thin}$ for \tar\ and other \tess\ planet host stars to be in the thick\ (TD) and thin disks\ (D). Figure \ref{toomre} shows the corresponding Toomre plot. We considered stars with $\rm P_{thick}/P_{thin}>10$ to be in the thick disk while stars in between\ ($\rm 0.1<P_{thick}/P_{thin}<10$) are ambiguous to judge. Up to now, \tess\ has detected five planet host stars located in the in-between region: TOI 118 \citep{Esposito2019}, TOI 144 \citep{Huang2018SE}, TOI 172 \citep{Rodriguez2019}, TOI 186 \citep{Trifonov2019,Dragomir2019} and TOI 197 \citep{Huber2019}. Table \ref{relative_probability} lists their relative probabilities and none of them show clear-cut thick-disk probability. However, we obtained a large relative probability\ ($\rm P_{thick}/P_{thin}$\ =\ 6482) for \tar, indicating it is very likely a thick-disk star. \cite{Soubiran2003} showed that thick-disk stars tend to have much lower metallicity than thin-disk stars. Therefore, our metallicity measurement $\rm [Fe/H] = -0.12 \pm 0.09$, based on the HARPS spectra, is consistent a thick-disk origin.

\begin{table}[htbp]
    \centering
    \caption{Relative probability for \tess\ stars with ambiguous separation between thick and thin components}
    \begin{tabular}{cc}
        \hline\hline
        Star       &$\rm P_{thick}/P_{thin}$\\\hline
        TOI-118              &4.825 \\
        TOI-144              &0.127 \\
        TOI-172              &1.430 \\
        TOI-186              &0.125 \\
        TOI-197              &0.292\\\hline
        LHS 1815             &6482\\
         \hline\hline
    \end{tabular}
    \label{relative_probability}
\end{table}

\par
\begin{figure}[htbp]
\centering
\includegraphics[width=9cm]{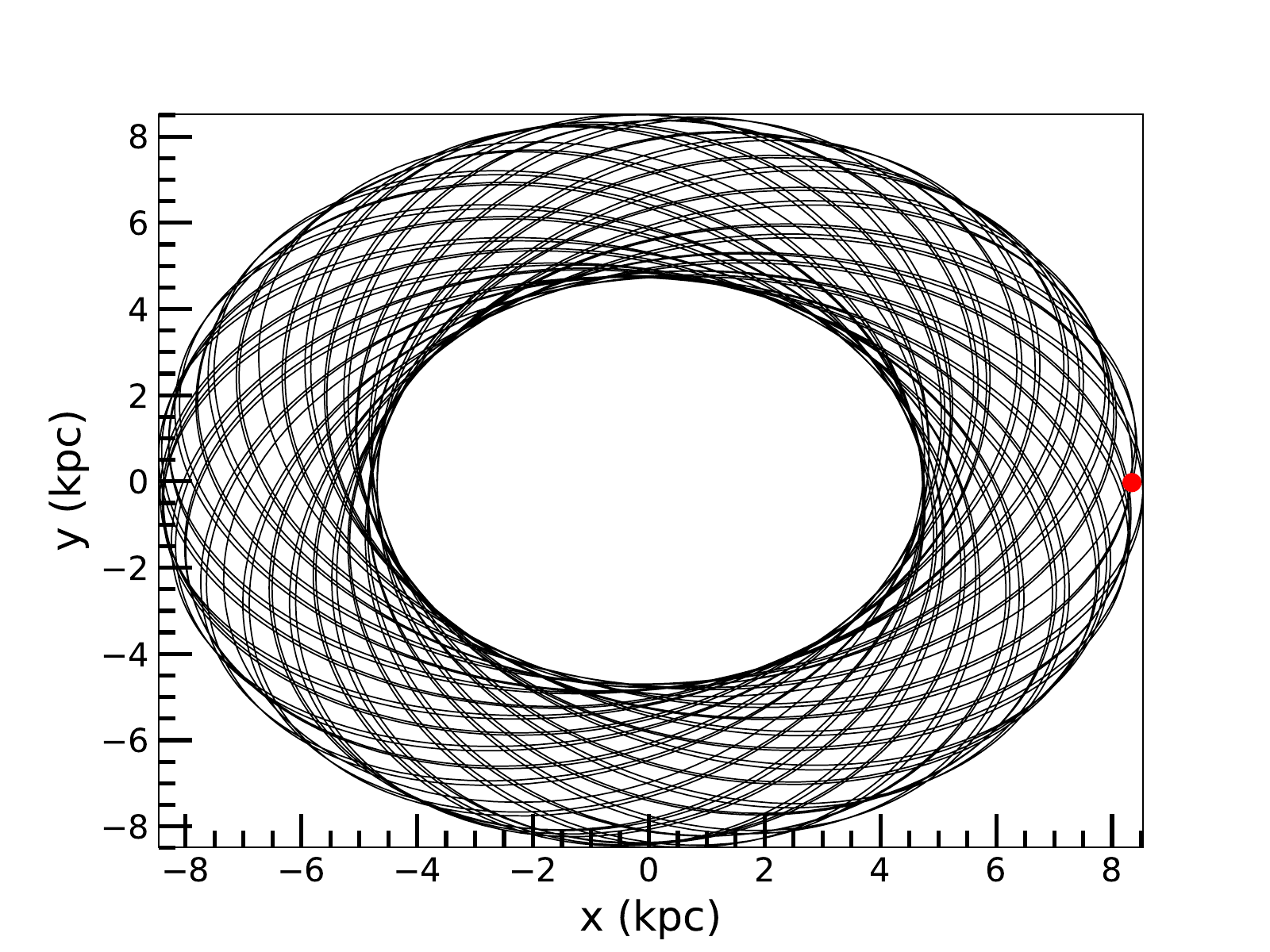}
\includegraphics[width=9cm]{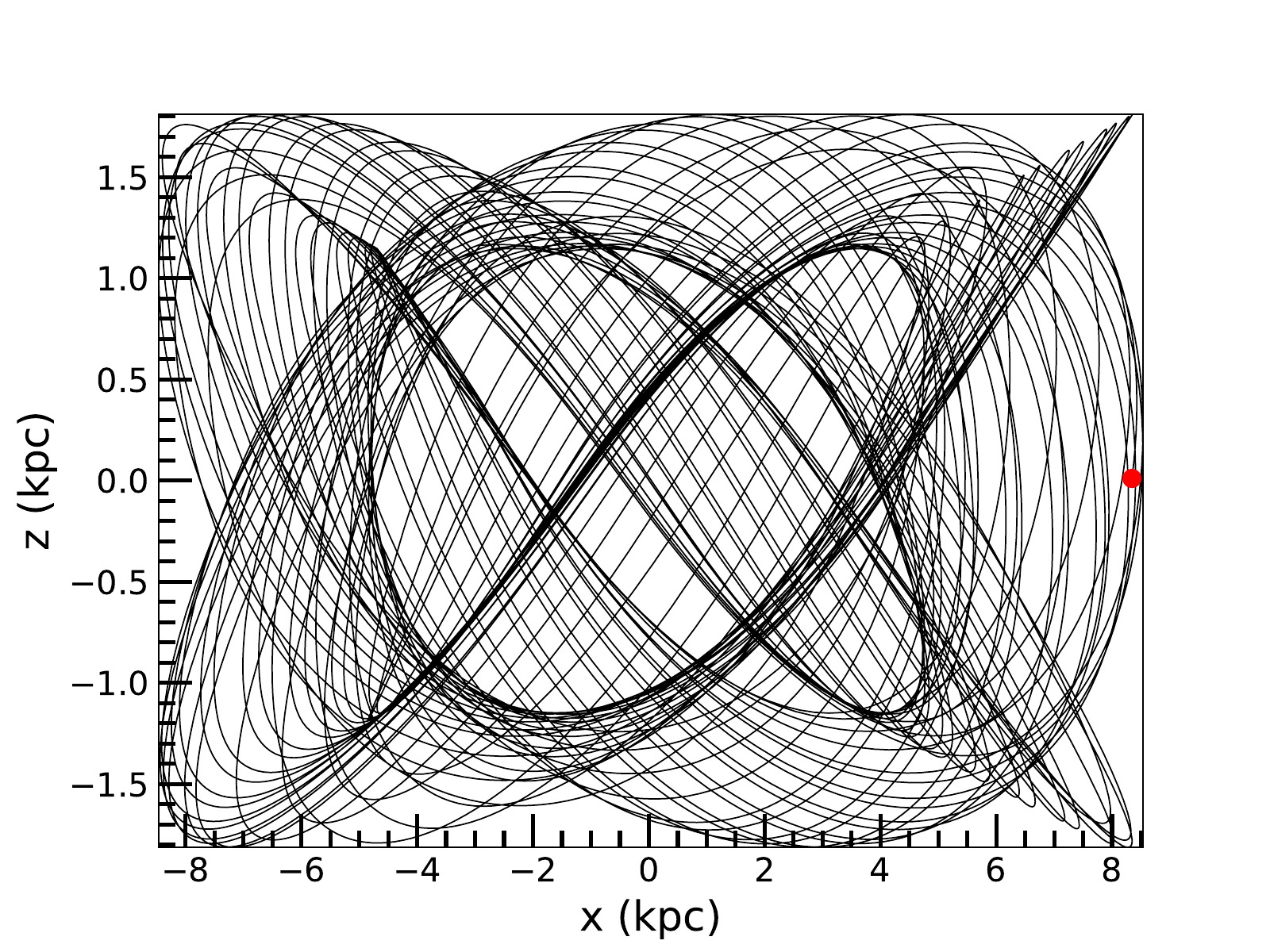}
\caption{{\it Top panel}: The orbit of \tar\ in the Galactic potential MWPotential2014 obtained using \textbf{galpy} \citep{Bovy2015} in the top down view. {\it Bottom panel}: The same orbit but viewed edge-on. The red dots represent the present position of the star.}
\label{orbital}
\end{figure}

\begin{figure}[htbp]
\centering
\includegraphics[width=9cm]{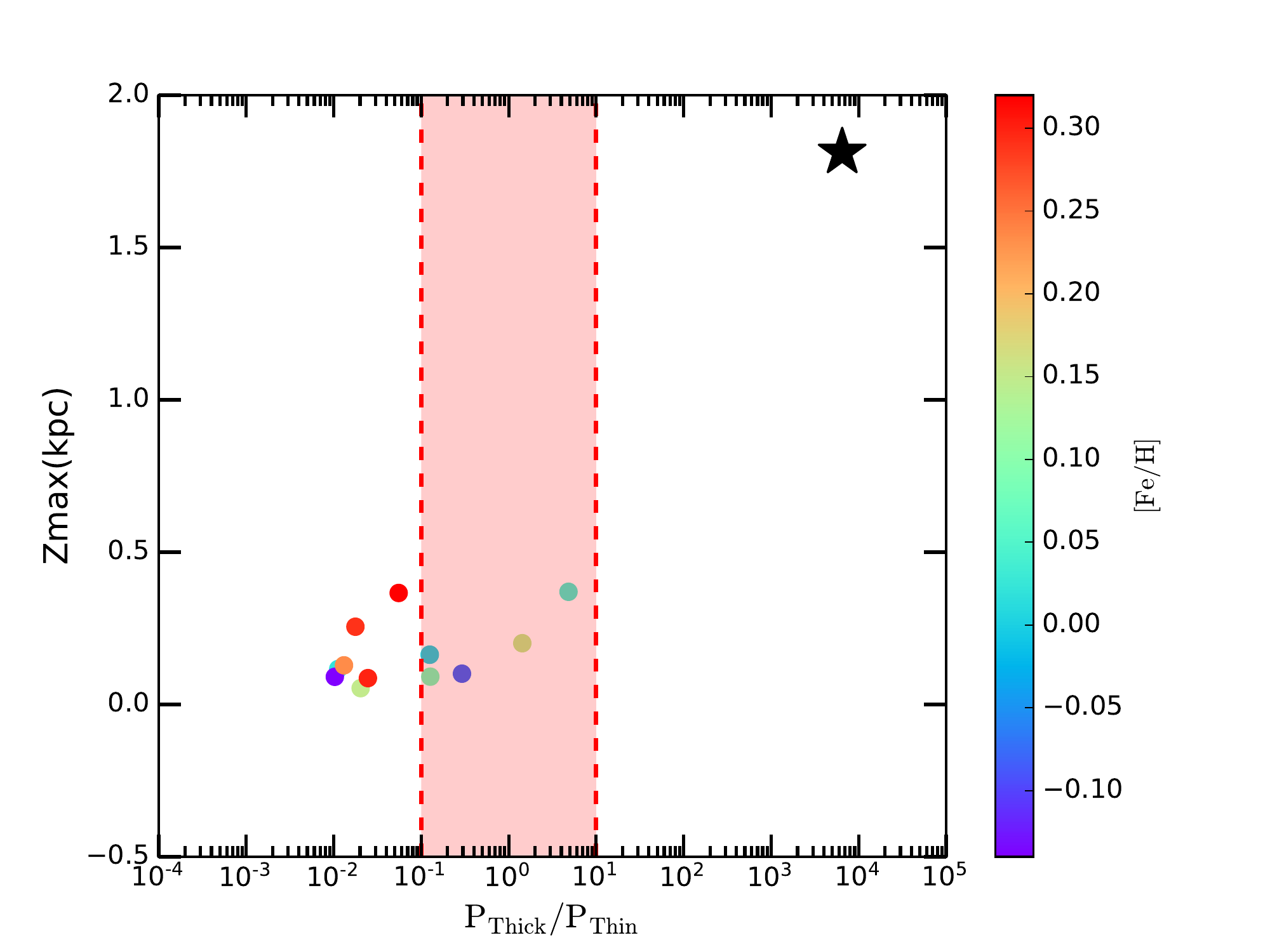}
\caption{$\rm Z_{max}$ vs. relative probability $\rm P_{thick}/P_{thin}$ for all \tess\ planets' host stars with \gaia\ radial velocity. $\rm Z_{max}$ is the expected maximal height of stars above the Galactic plane. Different colors represent different metallicities. The red vertical dashed lines mean relative probability\ =\ 0.1 and 10. \tar\ is marked as a black star at the top right.}
\label{planet_zmax_tess}
\end{figure}

 In order to gain insight into further dynamical information, we used {\textbf{galpy}} \citep{Bovy2015} to simulate the orbit of \tar. We initialized the orbit using RA, DEC, star distance, proper motions in two directions and heliocentric line-of-sight velocity. We integrated the orbit from t\ =\ 0 to t\ =\ 10 Gyr in a general potential: {\textbf{MWPotential2014}}, saving the orbit for 10000 steps. The orbital result of \tar\ is shown in Figure \ref{orbital}. The maximal height $\rm Z_{max}$ of \tar\ above the plane of the orbit is 1.8\ kpc, consistent with our thick disk conclusion before. For comparison, we plot $\rm Z_{max}$ and the relative probability of all \tess\ planet host stars in Figure \ref{planet_zmax_tess}. It is clear that the five TOI stars located in the region between the thin and thick disks are more likely to belong to the Galactic thin disk given their small $\rm Z_{max}$. \tar\ is moving upwards currently; an additional orbital integration analysis shows that \tar\ will spend $\rm \sim 14\ Myr$ to first reach 1\ kpc above the Galactic plane. Before \tar\ reaches the plane again, we have a probability about 33\% to see it\ ($\rm Z<1kpc$). 

\begin{table*}[htbp]
    \centering
    \caption{Final parameters of \tar b}
    \begin{tabular}{ccc}
        \hline\hline
        Parameter       &Value       &Prior\\\hline
        \it{Fitting parameters}\\
        $\rm P_{orb}$ (days)   &$3.81433\pm0.00003$  &$\mathcal{N}^{[1]}$ (3.814\ ,\ $0.1^{2}$)\\
        $\rm T_{C}$ (BJD)   &$2458327.4161\pm0.0016$ &$\mathcal{N}$ (2458327.4\ ,\ 0.1) \\
        $\rm R_{P}/R_{\star}$   &$0.0199\pm0.0009$    &$\mathcal{U}^{[2]}$ (0.005\ ,\ 0.05)\\
        $\rm a/R_{\star}$       &$17.403\pm2.816$       &$\mathcal{N}$ (16\ ,\ 3)\\
        $i$ (deg)                &$88.125\pm1.113$      &$\mathcal{U}$ (0\ ,\ 180)\\
        $\rm q_{1}$                &$0.26\pm0.19$       &$\mathcal{U}$ (0\ ,\ 1)\\
        $\rm q_{2}$                &$0.35\pm0.26$       &$\mathcal{U}$ (0\ ,\ 1)\\
        $K$ ($\rm m\ s^{-1}$)                    &$2.7\pm1.0$    
        &$\mathcal{U}$ (0\ ,\ 10)\\
        $\rm \gamma_{rel}$ ($\rm m\ s^{-1}$)      &$-0.38\pm0.64$    &$\mathcal{U}$ (-10\ ,\ 10)\\
        $\rm \sigma_{J}$  ($\rm m\ s^{-1}$)          &$2.0\pm0.6$    &$\mathcal{J}^{[3]}$ (0.1\ ,\ 10)\\
        $e$                     &0  &Fixed  \\
        $\omega$ (deg)          &90 &Fixed  \\
        \it{Derived parameters}\\
        $\rm R_{P}$ ($\rm R_{\oplus}$)   &$1.088\pm 0.064$   \\
        $\rm M_{P}$ ($\rm M_{\oplus}$)$^{[4]}$   &$4.2\pm1.5$ \\
        $\rm a$ (AU)    &$0.0404\pm0.0094$    \\
        $\rm T_{eq}$ (K)$^{[5]}$            &$617\pm84$             \\

         \hline\hline 
    \end{tabular}
    \begin{tablenotes}
    \item[1]  [1]\ $\mathcal{N}$($\mu\ ,\ \sigma$) means a normal prior with mean $\mu$ and standard deviation $\sigma$. 
    \item[2]  [2]\ $\mathcal{U}$(a\ , \ b) stands for a uniform prior ranging from a to b.
    \item[3]  [3]\ $\mathcal{J}$(a\ , \ b) stands for a Jeffrey's prior with the same limits.
    \item[4]  [4]\ This is not a statistically significant measurement. $3\sigma$ mass upper-limit is $\rm 8.7\ M_{\oplus}$. 
    \item[5]  [5]\ Suppose albedo = 0 and there is no heat distribution here. 
    \end{tablenotes}
    \label{planetparam}
\end{table*}

\section{Discussion and Conclusion}\label{dc}
\tar b is the first thick-disk planet detected by \tess. It has a radius of $\rm R_{P}=1.088\pm 0.064\ R_{\oplus}$ and a mass of $\rm M_{P}=4.2\pm1.5\ M_{\oplus}$. The proximity of \tar\ and its interesting kinematic features makes it a system worth further characterization. 

\subsection{Prospects on Future Follow-up Observations}

Given the brightness of \tar, it is an attractive target for precise RV measurements with high resolution spectroscopy facilities. Those will lead to precise mass measurement of the transiting planet and will be used to search for other planets in the system. A precise planet mass will give an improved estimate of the suitability of \tar b for atmospheric characterization. The rotation period of \tar\ is well separated from the orbital period of the planet, making it possible to smooth out the effect from stellar activity. 

\par In addition, since \tar\ is nearby\ ($29.87\pm0.02$ pc), future release of Gaia time series astrometry can be used to look for massive objects\ (massive planets and brown dwarfs) at wide orbits, with potential partial overlap with objects on orbits that radial velocities will be sensitive to.

\par To evaluate the feasibility of high-quality atmospheric characterization by \jwst\ \citep{Gardner2006}, 
we first use the Transmission Spectroscopy Metric\ (TSM) formulated by \cite{Kempton2018} and
we find $\rm TSM\sim2.5^{+3.8}_{-1.3}$ for \tar. \cite{Kempton2018} recommends that planets with $\rm TSM > 10$ for $\rm R_{p} < 1.5\ R_{\oplus}$ are high-quality atmospheric characterization targets. The relatively large TSM uncertainty due to the weak constraint on the planet mass results in unclear determination on whether \tar\ is a good\ (although unlikely the best) target for transmission spectroscopy studies. In addition, we compute the Emission Spectroscopy Metric\ (ESM) for \tar\ and we find $\rm ESM\sim1.9^{+1.0}_{-0.8}$. Given the recommended threshold $\rm ESM=7.5$ from \cite{Kempton2018}, \tar\ is not an ideal target for emission spectroscopy researches, either.




\subsection{Planet Formation Efficiency in Thin and Thick Disk?}
\par A followup statistical work about the planet formation efficiency in the thin and thick disk is ongoing\ (Gan et al., in prep) based on all \tess\ planet candidates detected in the Southern Hemisphere. The current \tess\ survey for the Northern Hemisphere will be an excellent opportunity to further examine this subject. First, \tess\ focuses on finding exoplanets around nearby bright stars and most TOIs have precise astrometry and RV measurement from \gaia\ DR2, which can determine their thin, thick and halo origin. Second, LAMOST\ (The Large Sky Area Multi-Object Fiber Spectroscopic Telescope, \citealt{Cui2012}) can provide chemical element abundance measurements to check the classification for a large number of stars. 
\par We emphasize that here we only consider the formation efficiency for nearby bright stars. Faint stars\ ($\rm G > 13$ mag) at relatively large distances may not have RV measurement 
from \gaia\ DR2, leading to a poor separation between thin and thick disks. Future surveys such as DESI \citep{Aghamousa2016} and spectroscopic observations from SPIRou \citep{Challita2018} shall remedy this situation. \\

\facilities{ TESS, ESO 3.6 m: HARPS, 4.1-m Southern Astrophysical Research (SOAR), Gemini-South, LCO:1.0m\ (Sinistro), WASP-south, Gaia}

\software{ AstroImageJ \citep{Collins2017}, TERRA \citep{Anglada2012}, SERVAL \citep{Zechmeister2018}, SpecMatch-Emp \citep{Yee2017}, lightkurve \citep{lightkurve}, EXONAIER \citep{Espinoza2016}, batman \citep{Kreidberg2015}, radvel \citep{Fulton2018}, emcee \citep{Foreman2013}}

\section{Acknowledgement}
We thank Sharon Xuesong Wang, Chao Liu and Weicheng Zang for their insights and advice.
Funding for the TESS mission is provided by NASA's Science Mission directorate. 
This work is partly supported by the National Science Foundation of China (Grant No. 11390372 and 11761131004 to SM and GTJ). 
We acknowledge the use of TESS Alert data from pipelines at the TESS Science Office and at the TESS Science Processing Operations Center. 
Resources supporting this work were provided by the NASA High-End Computing (HEC) Program through the NASA Advanced Supercomputing (NAS) Division at Ames Research Center for the production of the SPOC data products.
J.G.W. is supported by a grant from the John Templeton
Foundation. The opinions expressed in this publication are
those of the authors and do not necessarily reflect the views of
the John Templeton Foundation.
C.Z. is supported by a Dunlap Fellowship at the Dunlap Institute for Astronomy \& Astrophysics, funded through an endowment established by the Dunlap family and the University of Toronto.
Some of the observations in the paper made use of the High-Resolution Imaging instrument Zorro at Gemini-South). Zorro was funded by the NASA Exoplanet Exploration Program and built at the NASA Ames Research Center by Steve B. Howell, Nic Scott, Elliott P. Horch, and Emmett Quigley.
This research has made use of the Exoplanet Follow-up Observation Program website, which is operated by the California Institute of Technology, under contract with the National Aeronautics and Space Administration under the Exoplanet Exploration Program. 
This paper includes data collected by the TESS mission, which are publicly available from the Mikulski Archive for Space Telescopes\ (MAST). 
This research made use of observations from the LCO network, WASP-South and ESO: 3.6m\ (HARPS). 


\bibliographystyle{apj}
\bibliography{apj-jour,planet}
\end{document}